\documentclass[apj,iop,twocolappendix,numberedappendix]{emulateapj}
\usepackage[colorlinks,citecolor=blue,linkcolor=blue,urlcolor=blue,dvips]{hyperref}

\newcommand {\eg} {{\it e.g.}}

\newcommand {\be} {\begin{equation}}
\newcommand {\ee} {\end{equation}}
\newcommand {\bea} {\begin{eqnarray}}
\newcommand {\eea} {\end{eqnarray}}

\begin{document}

\title{Constraining the location of gamma-ray flares in luminous blazars}
\shorttitle{Locating gamma-ray flares in blazars}

\author{Krzysztof~Nalewajko\altaffilmark{1,2}, Mitchell C. Begelman\altaffilmark{1,3}, and Marek Sikora\altaffilmark{4}}
\shortauthors{Nalewajko, Begelman \& Sikora}

\altaffiltext{1}{JILA, University of Colorado and National Institute of Standards and Technology, 440 UCB, Boulder, CO 80309, USA; {\tt knalew@jila.colorado.edu}}
\altaffiltext{2}{NASA Einstein Postdoctoral Fellow}
\altaffiltext{3}{Department of Astrophysical and Planetary Sciences, University of Colorado, UCB 389, Boulder, CO 80309, USA}
\altaffiltext{4}{Nicolaus Copernicus Astronomical Center, Bartycka 18, 00-716 Warsaw, Poland}

\begin{abstract}
Locating the gamma-ray emission sites in blazar jets is a long-standing and highly controversial issue. We investigate jointly several constraints on the distance scale $r$ and Lorentz factor $\Gamma$ of the gamma-ray emitting regions in luminous blazars (primarily flat spectrum radio quasars, FSRQs). Working in the framework of one-zone external radiation Comptonization (ERC) models, we perform a parameter space study for several representative cases of actual gamma-ray flares in their multiwavelength context. We find a particularly useful combination of three constraints: from an upper limit on the collimation parameter $\Gamma\theta \lesssim 1$, from an upper limit on the synchrotron self-Compton (SSC) luminosity $L_{\rm SSC} \lesssim L_{\rm X}$, and from an upper limit on the efficient cooling photon energy $E_{\rm cool,obs} \lesssim 100\;{\rm MeV}$. These three constraints are particularly strong for sources with low accretion disk luminosity $L_{\rm d}$. The commonly used intrinsic pair-production opacity constraint on $\Gamma$ is usually much weaker than the SSC constraint. The SSC and cooling constraints provide a robust lower limit on the collimation parameter $\Gamma\theta \gtrsim 0.1 - 0.7$. Typical values of $r$ corresponding to moderate values of $\Gamma \sim 20$ are in the range $0.1 - 1\;{\rm pc}$, and are determined primarily by the observed variability time scale $t_{\rm var,obs}$. Alternative scenarios motivated by the observed gamma-ray/mm connection, in which gamma-ray flares of $t_{\rm var,obs} \sim$ a few days are located at $r \sim 10\;{\rm pc}$, are in conflict with both the SSC and cooling constraints. Moreover, we use a simple light travel time argument to point out that the gamma-ray/mm connection does not provide a significant constraint on the location of gamma-ray flares. We argue that spine-sheath models of the jet structure do not offer a plausible alternative to external radiation fields at large distances, however, an extended broad-line region is an idea worth exploring. We propose that the most definite additional constraint could be provided by determination of the synchrotron self-absorption (SSA) frequency for correlated synchrotron and gamma-ray flares.
\end{abstract}


\section{Introduction}

Blazars are a class of active galaxies, whose broad-band emission is dominated by non-thermal components produced in a relativistic jet pointing toward us \citep{1995PASP..107..803U}.
Due to the relativistic luminosity boost, many of these sources outshine their host galaxies by orders of magnitude, making them detectable at cosmological distances.
The brightest blazars, belonging to the subclasses known as flat-spectrum radio quasars (FSRQs) and low-synchrotron-peaked BL Lacertae objects (LBLs), radiate most of their energy in MeV/GeV gamma-rays \citep{1998MNRAS.299..433F}.
The origin of this gamma-ray emission has been debated for a long time, with proposed mechanisms including external-radiation Comptonization \citep[ERC;][]{1992A&A...256L..27D,1994ApJ...421..153S}, synchrotron self-Comptonization \citep[SSC;][]{1992ApJ...397L...5M,1996ApJ...461..657B}, and hadronic processes \citep[\eg,][]{1992A&A...253L..21M,2000NewA....5..377A,2001APh....15..121M}. The emerging consensus favors the ERC process \citep{1998MNRAS.301..451G,1999ApJ...527..132M,2001ApJ...553..683H,2009ApJ...704...38S,2013ApJ...768...54B}, especially for blazars with high-power jets \citep{2012ApJ...752L...4M}.

Several theoretical models have been proposed for energy dissipation and particle acceleration in relativistic blazar jets.
To discriminate among these models, it is crucial to pinpoint the location along the jet where the bulk of the non-thermal radiation is produced.
Several lines of argumentation have led blazar researchers to answers varying by almost 3 orders of magnitude.

Gamma-ray radiation at GeV energies can escape from the quasar environment, avoiding absorption, if it is produced at distances from the central engine $r \gtrsim 0.01\;{\rm pc}$ \citep{1996MNRAS.280...67G}.
At these smallest allowed distances, the dominant external radiation component in the jet co-moving frame is the direct emission of the accretion disk \citep[\eg,][]{2002ApJ...575..667D}.
At distances of $r \sim 0.1\;{\rm pc}$, the co-moving external radiation is dominated by broad emission lines (BEL) \citep[\eg,][]{1994ApJ...421..153S}.
For an emitting region propagating with a typical Lorentz factor of $\Gamma \simeq 20$, the observed variability time scale $\sim r/(\Gamma^2c)$ expected from radiation produced at such distances is several hours, which is consistent with the shortest variability time scales probed by the \emph{Fermi} Large Area Telescope (LAT) \citep{2010MNRAS.405L..94T,2013ApJ...766L..11S,2013A&A...557A..71R}. The likely dissipation mechanism at these distances depends on the efficiency of energy flux conversion from magnetic (Poynting flux) to inertial (kinetic energy flux) forms \citep{2005ApJ...625...72S}. In particle-dominated jets, internal shocks can operate with reasonable efficiency, provided that the jet acceleration mechanism is strongly modulated \citep{2001MNRAS.325.1559S}. In magnetically dominated jets, shocks are generally expected to be weak \citep[but see][]{2012MNRAS.422..326K}, however, in right circumstances the jet magnetic fields could be dissipated directly in the process of magnetic reconnection \citep[\eg,][]{2006A&A...450..887G,2009MNRAS.395L..29G}.

At distances of $r \gtrsim 1\;{\rm pc}$, external radiation fields are dominated by the infrared (IR) thermal radiation emitted by hot dust \citep{2000ApJ...545..107B}. The gamma-ray radiation produced at these distances is expected to vary over a few days. The associated synchrotron radiation should be transparent at wavelengths $\lambda_{\rm syn,obs} \lesssim 1\;{\rm mm}$, and in some sources a fairly good correlation was observed between the optical and millimeter signals \citep[\eg,][]{2008ApJ...675...71S}, or between the gamma-ray and millimeter signals \citep[\eg,][]{2012ApJ...758...72W}. At these distances, reconfinement shocks arising from the interaction of the jet with the external medium provide an alternative dissipation mechanism \citep[\eg,][]{2012MNRAS.420L..48N}.

The structure of blazar jets can be at least partially resolved with interferometric radio/mm observations.
Typically, it includes a stationary core and a succession of knots propagating superluminally downstream from the core.
The core could be a photosphere due to the synchrotron self-absorption process (certainly at wavelengths longer than $7\,{\rm mm}$), or an optically thin physical structure presumably resulting from reconfinement shocks \citep{2009ASPC..402..194M}. There is substantial evidence that many major gamma-ray flares in blazars are accompanied by radio/mm outbursts, and/or ejection (estimated moment of passing through the apparent position of the core) of superluminal radio/mm knots \citep[\eg,][]{2012arXiv1204.6707M}. While radio/mm outbursts are typically much longer ($\sim$ weeks/months) than gamma-ray flares ($\sim$ hours/days), the gamma-ray flares are often found between the onset and the peak of the mm outbursts \citep{2003ApJ...590...95L,2011A&A...532A.146L}. This gamma-ray/mm connection is used to argue for gamma-ray flares being produced at distance scales of $r \sim 10 - 20\;{\rm pc}$ (\eg, \citealt{2011ApJ...726L..13A,2011ApJ...735L..10A}; see also \citealt{2008ApJ...675...71S}). At these distances, the external radiation field is still likely dominated by thermal dust emission, although its energy density is expected to fall off rapidly with $r$. In order to explain short variability time scales of gamma-ray flares at such distances, very strong jet collimation is required.

In this work, we study the parameter space of location $r$ and Lorentz factor $\Gamma$ of the emitting regions responsible for major gamma-ray flares in luminous blazars.\footnote{In some blazar studies, multiple emitting regions were deemed necessary \citep[\eg,][]{2012ApJ...760...69N}. However, in any case where a coherent gamma-ray flare is observed, one can consider only the emitting region dominating the gamma-ray emission.}
We use 5 direct observables --- gamma-ray luminosity $L_\gamma$, gamma-ray variability time scale $t_{\rm var,obs}$, synchrotron luminosity $L_{\rm syn}$ (or the Compton dominance parameter $q = L_\gamma/L_{\rm syn}$), X-ray luminosity $L_{\rm X}$, and accretion disk luminosity $L_{\rm d}$ --- and a minimal number of assumptions --- in particular the Doppler-to-Lorentz factor ratio $\mathcal{D}/\Gamma$, and the external radiation sources covering factors $\xi_{\rm BLR},\xi_{\rm IR}$ --- to derive 4 constraints in the $(r,\Gamma)$ plane related to the following parameters --- collimation parameter $\Gamma\theta$, synchrotron self-Compton luminosity $L_{\rm SSC}$, observed ERC photon energy corresponding to efficient electron cooling threshold $E_{\rm cool,obs}$, and observed ERC photon energy corresponding to intrinsic pair-production absorption threshold $E_{\rm max,obs}$ --- and 2 predictions for the following parameters --- synchrotron self-absorption characteristic observed wavelength $\lambda_{\rm SSA,obs}$, and minimum required jet power $L_{\rm j,min}$.
These constraints are then applied in several case studies of actual gamma-ray flares of prominent blazars for which detailed multiwavelength data are available, and for which all 5 observables can be securely estimated. 
Most of these cases have already been discussed in the literature, but here they are systematically and critically compared for the first time.

We begin by deriving our constraints in Section \ref{sec_constr}, followed by additional predictions in Section \ref{sec_pred}.
Then we present the case studies in Section \ref{sec_cases}.
We consider the sensitivity of our constraints to the most uncertain parameters in Section \ref{sec_sens}.
Our results are discussed in Section \ref{sec_disc} and summarized in Section \ref{sec_conc}.

\section{Constraints on $\lowercase{r}$ and $\Gamma$}
\label{sec_constr}

We consider an emitting region located at distance $r$ from the central supermassive black hole (SMBH), propagating with velocity $\beta=v/c$ and Lorentz factor $\Gamma=(1-\beta^2)^{-1/2}$. Parameters measured in the co-moving frame of the emitting region will be denoted with a prime. We should stress here that the Lorentz factor of the emitting region $\Gamma$ does not need to coincide with the jet Lorentz factor $\Gamma_{\rm j}$. While simple models explicitly assume that $\Gamma \simeq \Gamma_{\rm j}$, in some scenarios a significant difference between these values is inferred, e.g., in the spine-sheath model \citep{2005A&A...432..401G}, and in the minijet model \citep{2009MNRAS.395L..29G}.

For an observer located at viewing angle $\theta_{\rm obs}$ with respect to the emitting region velocity vector, the Doppler factor of the observed radiation is $\mathcal{D}=[\Gamma(1-\beta\cos\theta_{\rm obs})]^{-1}$.
In blazars, the value of $\mathcal{D}$ is of the same order as $\Gamma$, but the actual ratio $\mathcal{D}/\Gamma$ is a major source of uncertainty in constraining $r$ and $\Gamma$.
In the case of a very compact emitting region, for $\theta_{\rm obs} \simeq 1/\Gamma$ we have $\mathcal{D}/\Gamma \simeq 1$, and for $\theta_{\rm obs} \simeq 0$ we have $\mathcal{D}/\Gamma \simeq 2$.
However, in a conical jet, elements of the emitting region may span a significant range of $\theta_{\rm obs}$, and thus a significant range of $\mathcal{D}/\Gamma$.
The effective value of $\mathcal{D}/\Gamma$ depends not only on the mean $\theta_{\rm obs}$ of the emitting region, but also on its opening angle $\theta$.
In particular, for emitting regions with $\Gamma\theta \sim 1$, we expect that $\mathcal{D}/\Gamma \lesssim 1$.

The values of $\Gamma_{\rm j}$ and $\mathcal{D}$ for individual sources can be evaluated independently by analyzing the radio structure of jets observed with VLBI techniques \citep{2005AJ....130.1418J}, and many such results are available for the MOJAVE sample \citep{2009A&A...494..527H}.
Therefore, it is now possible to make an informed choice of $\mathcal{D}/\Gamma_{\rm j}$ for many studied sources. However, as we will discuss later, this does not work equally well for all sources.
In this work, we decided to adopt $\mathcal{D}/\Gamma = 1$ for all analyzed sources, and we evaluate the effect of varying the value of $\mathcal{D}/\Gamma$ in Section \ref{sec_sens}.

\subsection{Collimation constraint}
\label{sec_constr_coll}

We assume that the emitting region has characteristic size $R$, which is related to the co-moving variability time scale via $R \simeq ct_{\rm var}'$.
The variability time scale scales like $t'_{\rm var} = \mathcal{D}t_{\rm var,obs}/(1+z)$, where $z$ is the blazar redshift.
The most reliable estimate of the observed variability time scale $t_{\rm var,obs}$ is the flux-doubling time scale measured with respect to the flare peak.
We can also relate $R$ to the location of the emitting region via $R \simeq \theta r$.
Again, we distinguish $\theta$ from the jet opening angle $\theta_{\rm j}$, demanding only that $\theta \le \theta_{\rm j}$. It is convenient to combine $\theta$ with the Lorentz factor $\Gamma$ to define \emph{the collimation parameter} $\Gamma\theta$.
We can now write the source Lorentz factor as a function of $\Gamma\theta$:
\be
\label{eq_constr_Gammatheta}
\Gamma(r,\Gamma\theta) \simeq \left(\frac{\mathcal{D}}{\Gamma}\right)^{-1/2}\left[\frac{(1+z)(\Gamma\theta)r}{ct_{\rm var,obs}}\right]^{1/2} \,.
\ee
There are strong observational and theoretical indications that $\Gamma_{\rm j}\theta_{\rm j} < 1$ for blazar jets.
Jet opening angle at the scale of tens of pc was measured in a substantial sample of blazars using VLBI imaging, with the typical result of $\Gamma_{\rm j}\theta_{\rm j} \sim 0.1-0.2$ \citep{2009A&A...507L..33P,2013A&A...558A.144C}.
Numerical simulations of acceleration and collimation of external-pressure-supported relativistic jets also find that after the acceleration is complete, $\Gamma_{\rm j}\theta_{\rm j}\lesssim 1$ \citep{2009MNRAS.394.1182K,2010NewA...15..749T}.
However, the relation between the collimation parameter of the jet $\Gamma_{\rm j}\theta_{\rm j}$ and the collimation parameter of the emitting region $\Gamma\theta$ is unclear.
On one hand, we expect that $\theta \le \theta_{\rm j}$, on the other hand, it is possible that $\Gamma > \Gamma_{\rm j}$.
Therefore, here we adopt a relatively conservative \emph{collimation constraint}, defined as $\Gamma\theta \lesssim 1$.

\subsection{SSC constraint}
\label{sec_constr_ssc}

We assume that the gamma-ray emission is produced by Comptonization of external radiation (ERC) by a population of ultrarelativistic electrons, and that the apparent gamma-ray luminosity $L_\gamma$ (hereafter understood as the peak of $\nu L_{\gamma,\nu}$ SED, as opposed to the bolometric luminosity $L_{\rm\gamma,bol} = \int L_{\gamma,\nu}\;{\rm d}\nu$) measured by \emph{Fermi}/LAT represents $L_{\rm ERC}$, the peak luminosity of the ERC component.
The same electrons produce the synchrotron and the SSC components, of which at least the former should contribute to the observed spectral energy distributions (SEDs) as indicated by fast optical/IR flares often correlated with the gamma rays.
The three luminosities  --- $L_{\rm ERC}$, $L_{\rm syn}$ and $L_{\rm SSC}$ --- can be related to the co-moving energy densities of external radiation $u_{\rm ext}'$, magnetic fields $u_{\rm B}' = B'^2/(8\pi)$, and synchrotron radiation $u_{\rm syn}' \simeq L_{\rm syn}/(4\pi c\mathcal{D}^4R^2)$, respectively.
On one hand, we have $L_{\rm SSC}/L_{\rm syn} \simeq g_{\rm SSC}(u_{\rm syn}'/u_{\rm B}')$, where $g_{\rm SSC} = (L_{\rm SSC}/L_{\rm syn})/(L_{\rm SSC,bol}/L_{\rm syn,bol}) \simeq 3/4$ is a bolometric correction factor (mainly due to spectral shape and source geometry).
On the other hand, we can define a Compton dominance parameter
\be
\label{eq_q}
q = \frac{L_\gamma}{L_{\rm syn}} \simeq g_{\rm ERC}\left(\frac{\mathcal{D}}{\Gamma}\right)^2\left(\frac{u_{\rm ext}'}{u_{\rm B}'}\right)\,,
\ee
where $g_{\rm ERC} = (L_{\rm ERC}/L_{\rm syn})/(L_{\rm ERC,bol}/L_{\rm syn,bol}) \simeq 1/2$ is a bolometric correction factor (mainly due to Klein-Nishina effects), and the $(\mathcal{D}/\Gamma)^2$ factor reflects the beaming profile of the ERC component in the case of flat $\nu L_\nu$ SED \citep{1995ApJ...446L..63D}.
The co-moving energy density of external radiation is related to the accretion disk luminosity $L_{\rm d}$ via
\be
\label{eq_uext}
u_{\rm ext}' \simeq \frac{\zeta(r)\Gamma^2L_{\rm d}}{3\pi cr^2}\,.
\ee
Here, $\zeta(r)$ is a function that describes the composition of external radiation fields, including contributions from the broad-line region (BLR), the dusty torus producing infrared emission (IR), and the direct accretion disk radiation:
\bea
\label{eq_zeta}
\zeta(r) &\simeq& \frac{0.4\xi_{\rm BLR}(r/r_{\rm BLR})^2}{1+(r/r_{\rm BLR})^4} +\frac{0.4\xi_{\rm IR}(r/r_{\rm IR})^2}{1+(r/r_{\rm IR})^4}
+\frac{0.21R_{\rm g}}{r}\,,
\eea
where $\xi_{\rm BLR}$ is the covering factor of the BLR of characteristic radius $r_{\rm BLR}$, $\xi_{\rm IR}$ and $r_{\rm IR}$ are the analogous parameters of the dusty torus, and $R_{\rm g}$ is the gravitational radius of the SMBH (we explain the origin of this function in Appendix \ref{sec_ext_rad}).
In this work, we adopt the following scaling laws: $r_{\rm BLR} \simeq 0.1L_{\rm d,46}^{1/2}\;{\rm pc}$, and $r_{\rm IR} \simeq 2.5L_{\rm d,46}^{1/2}\;{\rm pc}$, where $L_{\rm d,46} = L_{\rm d}/(10^{46}\;{\rm erg\,s^{-1}})$ \citep{2009ApJ...704...38S}.
Putting the above relations together, we obtain a constraint on $\Gamma$:
\bea
\label{eq_constr_LSSC}
\Gamma(r,L_{\rm SSC}) &\simeq& \left[3\left(\frac{g_{\rm SSC}}{g_{\rm ERC}}\right)\left(\frac{L_{\rm syn}}{L_{\rm SSC}}\right)\left(\frac{L_\gamma}{\zeta(r)L_{\rm d}}\right)\right]^{1/8}
\nonumber\\&&
\times\left(\frac{\mathcal{D}}{\Gamma}\right)^{-1}\left[\frac{(1+z)r}{2ct_{\rm var,obs}}\right]^{1/4}\,.
\eea

The SSC component in the SEDs of luminous blazars peaks at the observed photon energy of $E_{\rm SSC,obs} \simeq 20\;{\rm neV}\times \mathcal{D}B_0' \gamma_{\rm peak}^4 /(1+z)$, where $B_0' = B'/(1\;{\rm G})$ and $\gamma_{\rm peak}$ is the characteristic random Lorentz factor of electrons contributing to the SED peaks.
We can estimate $\gamma_{\rm peak}$ from the observed photon energy of the SED peak of the ERC component $E_{\rm ERC,obs} \simeq \mathcal{D}\Gamma\gamma_{\rm peak}^2E_{\rm ext}(r)/(1+z)$,
where $E_{\rm ext}(r)$ is the energy of external radiation photons.
In order to account for the transition between the BLR and IR external radiation fields, we use the following approximation (see Appendix \ref{sec_ext_rad}):
\be
\label{eq_Eext}
E_{\rm ext}(r) \simeq \frac{E_{\rm BLR}}{1+(r/r_{\rm BLR})^3} + \frac{E_{\rm IR}}{1+(r/r_{\rm IR})^3}\,,
\ee
where $E_{\rm BLR} \simeq 10\;{\rm eV}$ and $E_{\rm IR} \simeq 0.3\;{\rm eV}$.
The magnetic field strength can be found from Equations (\ref{eq_q} -- \ref{eq_uext}):
\be
B' \simeq \frac{\mathcal{D}}{r}\left[\frac{8g_{\rm ERC}\zeta(r)L_{\rm d}}{3qc}\right]^{1/2}\,.
\ee
Combining the above formulas, we find:
%
\bea
E_{\rm SSC,obs}
&\simeq&
\frac{20\;{\rm neV}}{r}\frac{(1+z)}{\Gamma^2}\left[\frac{E_{\rm ERC,obs}}{E_{\rm ext}(r)}\right]^2
\times\nonumber\\&&
\left[\frac{8g_{\rm ERC}\zeta(r)L_{\rm d}}{3qc}\right]^{1/2}\,.
\eea
One can see that $E_{\rm SSC,obs}$ is a sensitive function of $E_{\rm ERC,obs}$ and $\Gamma$.
However, for $\Gamma = 20$, $E_{\rm ERC,obs} = 100\;{\rm MeV}$, $r = 1\;{\rm pc}$, $E_{\rm ext} = 1\;{\rm eV}$, $\zeta = 0.1$, $L_{\rm d} = 3\times 10^{45}\;{\rm erg\,s^{-1}}$, and $q = 10$, we find $E_{\rm SSC,obs} \simeq 6(1+z)\;{\rm keV}$.
Because SSC spectral components are very broad, in most cases they should peak around, or contribute significantly to, the soft/hard X-ray band.
Some blazars show spectral softening in the soft X-ray part of their SEDs, which was interpreted as a signature of the SSC component \citep{2011MNRAS.410..368B}.
However, in many sources the observed X-ray emission is harder than it would be if it were dominated by the SSC component \citep{2009ApJ...704...38S}.
Also, the observed X-ray variability is usually not well correlated with variability in the gamma-ray and optical bands \citep{2012ApJ...754..114H}.
In the case that the SSC component dominates the X-ray emission, we would expect that X-ray variability should be stronger than the optical/IR variability.
For example, in a simple scenario of varying number of energetic electrons at constant magnetic field we have $L_{\rm SSC} \propto L_{\rm syn}^2$.
As this is not the case for luminous blazars, we can only use the observed X-ray luminosity as an upper limit for the SSC luminosity \citep{2010ApJ...721.1383A}.
Therefore, our \emph{SSC constraint} is defined as $L_{\rm SSC} \lesssim L_{\rm X}$.

\subsection{Cooling constraint}

Rapid gamma-ray variability of blazars, with roughly time-symmetric light curve peaks, and tight energetic requirements for the brightest observed gamma-ray flares, indicate very efficient cooling of the underlying ultrarelativistic electrons.
The radiative cooling of electrons in luminous blazars is dominated by the ERC process with cooling time scale $t_{\rm cool}'(\gamma) \simeq 3m_{\rm e}c/(4\sigma_{\rm T}\gamma u_{\rm ext}')$, where $\gamma$ is the electron random Lorentz factor.
In general, $t_{\rm cool}'(\gamma)$ should be compared with the variability time scale $t_{\rm var}'$ (which is associated with the observed flux doubling timescale, see Section \ref{sec_constr_coll}), and adiabatic cooling time scale $t_{\rm ad}'$.\footnote{The adiabatic loss time scale is $t_{\rm ad}' \simeq r/(A\Gamma c) \simeq t_{\rm var}'/(A\Gamma\theta)$, where $A \le 1$. Therefore, as long as the collimation constraint $\Gamma\theta \lesssim 1$ is satisfied, we have $t_{\rm ad}' \gtrsim t_{\rm var}'$.}
Observations of roughly time-symmetric flares indicate that the cooling time scales do not exceed the observed flux decaying time scales, i.e., that $t_{\rm cool}'(\gamma) \lesssim t_{\rm var}'$.\footnote{Alternatively, the time-symmetric gamma-ray flares may indicate that the velocity vector of the emitting region is rapidly swinging relative to the line-of-sight. In such case, both the flux rise and decay time scales would be determined primarily by variations in the Doppler factor.}
We calculate a characteristic electron Lorentz factor $\gamma_{\rm cool}$ such that $t_{\rm cool}'(\gamma_{\rm cool}) \simeq t_{\rm var}'$,\footnote{This is different from a \emph{cooling break} which is obtained by equating the radiative and adiabatic energy loss rates.}
and a corresponding observed ERC photon energy $E_{\rm cool,obs} \simeq \mathcal{D}\Gamma\gamma_{\rm cool}^2E_{\rm ext}(r)/(1+z)$.
Taking the above together, we obtain the following constraint on $\Gamma$:
\bea
\label{eq_constr_Ecool}
\Gamma(r,E_{\rm cool,obs}) &\simeq& \left(\frac{\mathcal{D}}{\Gamma}\right)^{-1/4}\left[\frac{9\pi m_{\rm e}c^2r^2}{4\sigma_{\rm T}\zeta(r)L_{\rm d}t_{\rm var,obs}}\right]^{1/2}
\nonumber\\&&\times
\left[\frac{(1+z)E_{\rm ext}(r)}{E_{\rm cool,obs}}\right]^{1/4}\,.
\eea
Since the gamma-ray light curves based on the \emph{Fermi}/LAT data are typically calculated for photon energies $E > 100\;{\rm MeV}$, our \emph{cooling constraint} is defined as $E_{\rm cool,obs} \lesssim 100\;{\rm MeV}$.

Alternatively, the cooling time scale as a function of photon energy potentially can be estimated directly from gamma-ray observations, but this is only feasible for the very brightest events \citep{2012ApJ...758L..15D}.

\subsection{Internal gamma-ray opacity constraint}

The maximum observed gamma-ray photon energy $E_{\rm max,obs}$ is constrained at least by the pair-production absorption process due to soft radiation produced in the same emitting region \citep[\eg,][]{1995MNRAS.273..583D}. The peak cross section for the pair-production process is $\sigma_{\rm \gamma\gamma} \simeq \sigma_{\rm T}/5$ for soft photons of co-moving energy $E_{\rm soft}' \simeq 3.6(m_{\rm e}c^2)^2/E_{\rm max}'$. In the observer frame, the soft photon energy is
\bea
\label{eq_Esoftobs}
E_{\rm soft,obs} &\simeq& \frac{3.6(m_{\rm e}c^2)^2\mathcal{D}^2}{(1+z)^2E_{\rm max,obs}}
\nonumber\\
&\simeq& \frac{38\;{\rm keV}}{(1+z)^2}\left(\frac{E_{\rm max,obs}}{10\;{\rm GeV}}\right)^{-1}\left(\frac{\mathcal{D}}{20}\right)^2\,.
\eea
The optical depth for gamma-ray photons is:
\be
\label{eq_taugg1}
\tau_{\rm\gamma\gamma} = \sigma_{\rm\gamma\gamma}n_{\rm soft}'R \simeq \frac{(1+z)^2\sigma_{\rm T}L_{\rm soft}E_{\rm max,obs}}{72\pi(m_{\rm e}c^2)^2c^2\mathcal{D}^6t_{\rm var,obs}}\,.
\ee
As the observed soft photon energy $E_{\rm soft,obs}$ may fall outside any observed energy range, we relate the target soft radiation luminosity to the observed X-ray luminosity via a spectral index $\alpha$ such that
\be
L_{\rm soft} = L_{\rm X}\left(\frac{E_{\rm soft,obs}}{E_{\rm X}}\right)^{1-\alpha} \simeq \frac{[3.6(m_{\rm e}c^2)^2]^{1-\alpha}\mathcal{D}^{2-2\alpha}L_{\rm X}}{(1+z)^{2-2\alpha}E_{\rm X}^{1-\alpha}E_{\rm max,obs}^{1-\alpha}}\,.
\ee
Substituting this into Equation (\ref{eq_taugg1}), we obtain:
\be
\tau_{\rm\gamma\gamma} \simeq \frac{(1+z)^{2\alpha}\sigma_{\rm T}L_{\rm X}E_{\rm max,obs}^{\alpha}}{20\pi[3.6(m_{\rm e}c^2)^2]^{\alpha}c^2\mathcal{D}^{4+2\alpha}E_{\rm X}^{1-\alpha}t_{\rm var,obs}}\,.
\ee
For gamma-ray observations of blazars, it is typical to associate $E_{\rm max,obs}$ with $\tau_{\rm\gamma\gamma} \simeq 1$. This leads to the following constraint on $\Gamma$:
\bea
\label{eq_constr_Emax}
\Gamma(r,E_{\rm max,obs}) &\simeq& \left\{\frac{(1+z)^{2\alpha}\sigma_{\rm T}L_{\rm X}E_{\rm max,obs}^{\alpha}}{20\pi[3.6(m_{\rm e}c^2)^2]^{\alpha}c^2E_{\rm X}^{1-\alpha}t_{\rm var,obs}}\right\}^{\frac{1}{4+2\alpha}}
\nonumber\\&&
\times\left(\frac{\mathcal{D}}{\Gamma}\right)^{-1}\,.
\eea
In Section \ref{sec_cases}, we will demonstrate that the \emph{internal gamma-ray opacity constraint} is relatively weak compared to the SSC constraint.

Additional potential source of gamma-ray opacity is from the broad emission lines. To the first order of approximation, this would affect photons of observed energy:
\be
E_{\rm max,BLR,obs} \simeq \frac{3.6(m_{\rm e}c^2)^2}{(1+z)E_{\rm BLR}} \simeq \frac{94\;{\rm GeV}}{(1+z)}\left(\frac{E_{\rm BLR}}{10\;{\rm eV}}\right)^{-1}\,,
\ee
with the peak optical depth of
\bea
\tau_{\rm\gamma\gamma,BLR}(r)
&\simeq&
\frac{\xi_{\rm BLR}\sigma_{\rm T}L_{\rm d}(r_{\rm BLR}-r)}{20\pi c r_{\rm BLR}^2E_{\rm BLR}}
\simeq
71\left(\frac{\xi_{\rm BLR}}{0.1}\right)
\times\nonumber\\&&
\left(\frac{L_{\rm d}}{10^{46}\;{\rm erg\,s^{-1}}}\right)^{1/2}\left(\frac{r_{\rm BLR}-r}{r_{\rm BLR}}\right)\left(\frac{E_{\rm BLR}}{10\;{\rm eV}}\right)^{-1}
\eea
(again using the scaling $r_{\rm BLR} \simeq 0.1L_{\rm d,46}^{1/2}\;{\rm pc}$).
The high value of the peak optical depth indicates that absorption should become noticeable already at the threshold observed energy of $(m_{\rm e}c^2)^2/[(1+z)E_{\rm BLR}] \simeq 26\;{\rm GeV}/(1+z)/(E_{\rm BLR}/10\;{\rm eV})$.\footnote{Considering the ionized Helium lines with $E_{\rm BLR} \simeq 54\;{\rm eV}$, the threshold observed energy would shift to $\simeq 4.8\;{\rm GeV}/(1+z)$ \citep{2010ApJ...717L.118P}, however, this is only relevant for distance scales $r \ll r_{\rm BLR}$ that are not of interest here.} The actual strength of the BLR absorption features depends significantly on the BLR geometry, and determining it requires detailed calculations \citep[\eg,][]{2003APh....18..377D,2007ApJ...665.1023R,2012arXiv1209.2291T}.
We will briefly comment on the expected significance of the BLR absorption in those cases from Section \ref{sec_cases}, which allow for the emitting region to be located within $r_{\rm BLR}$.

\section{Predictions for given $\lowercase{r}$ and $\Gamma$}
\label{sec_pred}

\subsection{Synchrotron self-absorption}

Synchrotron radiation is subject to synchrotron self-absorption (SSA) process, which can produce a sharp spectral break. This is a powerful probe of the intrinsic radius of the source of synchrotron emission \citep[\eg,][]{2008ApJ...675...71S,2013ApJ...772...78B}. In the co-moving frame, the SSA break is expected at:
\be
\nu_{\rm SSA}' \simeq \frac{1}{3}\left(\frac{eB'}{m_{\rm e}^3c}\right)^{1/7}\frac{L_{\rm syn}'^{2/7}}{R^{4/7}}\,,
\ee
where we approximated the synchrotron luminosity at $\nu_{\rm SSA}'$ with the synchrotron energy distribution peak luminosity $L_{\rm syn}'$ (i.e., we assumed a flat synchrotron SED in the mid-IR/mm band; in any case $\nu_{\rm SSA}'$ depends only weakly on the spectral index of unabsorbed synchrotron emission).
Substituting relevant relations from previous sections, we find a constraint on $\Gamma$:
\bea
\label{eq_constr_nuabsobs}
\Gamma(r,\nu_{\rm SSA,obs}) &\simeq& \left[\frac{8g_{\rm ERC}e^2\zeta(r)L_{\rm d}L_\gamma^4}{3^{15}q^5m_{\rm e}^6c^{11}(1+z)^6\nu_{\rm SSA,obs}^{14}t_{\rm var,obs}^8r^2}\right]^{1/8}
\nonumber\\
&&\times\left(\frac{\mathcal{D}}{\Gamma}\right)^{-1}\,.
\eea
In luminous blazars, the SSA spectral break is typically observed in the sub-mm/radio band.
As the synchrotron radiation observed in this band probes lower electron energies than the $\sim {\rm GeV}$ gamma-ray radiation, a connection between these bands should be verified by studying variability correlations.
These are very challenging observations, and for most cases studied in Section \ref{sec_cases} such data are not available.
Therefore, in this work the \emph{SSA constraint} is limited to provide a prediction of what $\nu_{\rm SSA,obs}$ should be for each studied case.

\subsection{Jet energetics}

We can constrain the energy content of blazar jets underlying the observed gamma-ray flares by estimating two of its essential ingredients: the radiation energy density dominated by the gamma rays $u_\gamma'$, and the magnetic energy density $u_{\rm B}'$.
Because the production of gamma-ray radiation through the ERC process is very efficient, $u_\gamma'$ closely probes the high-energy end of the electron energy distribution.
Additional jet energy may be carried by cold/warm electrons and protons, the contribution of which is very uncertain.
For example, the number of cold electrons can be constrained by modeling the broad-band SEDs, but the low-energy electron distribution index is usually one of the most uncertain parameters.
On the other hand, the energy content of protons in blazar jets can be constrained only indirectly, by combining arguments such as interpretation of (hard) X-ray spectra of luminous blazars, and energetic coupling between the protons and electrons \citep{2011IAUS..275...59S}.
Rather than introducing extra parameters with highly uncertain values, we choose to discuss a firm lower limit $L_{\rm j,min}$ on the jet power required to produce the observed gamma-ray flares of blazars together with their synchrotron and SSC counterparts.

The radiation energy density can be written as:
\be
u'_\gamma \simeq \frac{L_\gamma}{4\pi c\mathcal{D}^4R^2} \simeq \left(\frac{\mathcal{D}}{\Gamma}\right)^{-6}\frac{(1+z)^2L_\gamma}{4\pi c^3\Gamma^6t_{\rm var,obs}^2}\,.
\ee
The magnetic energy density $u_{\rm B}'$ can be derived from the synchrotron luminosity $L_{\rm syn}$, which is related to the gamma-ray luminosity $L_\gamma$ through the Compton dominance parameter $q = L_\gamma/L_{\rm syn}$:
\be
u_{\rm B}' \simeq \left(\frac{\mathcal{D}}{\Gamma}\right)^2\frac{g_{\rm ERC}u_{\rm ext}'}{q} \simeq \left(\frac{\mathcal{D}}{\Gamma}\right)^2\frac{g_{\rm ERC}\zeta(r)\Gamma^2L_{\rm d}}{3\pi cqr^2}\,.
\ee
Instead of using these two energy densities separately, we will analyze their more useful combinations: their ratio and their sum. The ratio of the two energy densities is a measure of energy equipartition between the magnetic fields and the ultra-relativistic electrons. One can show that \citep[\emph{cf.}][]{2009ApJ...704...38S}:
\be
\label{eq_constr_ugammauB}
\frac{u_\gamma'}{u_{\rm B}'} \simeq \frac{L_\gamma L_{\rm SSC}}{g_{\rm SSC}L_{\rm syn}^2}\,,
\ee
therefore, this energy density ratio is proportional to $L_{\rm SSC}$, and it follows the same dependence on $r$ and $\Gamma$. The sum of the two energy densities constitutes a lower limit on the jet energy density $u_{\rm j,min}' = u_\gamma' + u_{\rm B}'$. The corresponding minimum jet power is given by $L_{\rm j,min} \simeq \pi c\Gamma^2R^2u'_{\rm j,min}$. Therefore, we can write $L_{\rm j,min} = L_{\rm j,\gamma,min} + L_{\rm j,B,min}$, where
\bea
\label{eq_constr_Ljmin1}
L_{\rm j,\gamma,min} &=& \left(\frac{\mathcal{D}}{\Gamma}\right)^{-4}\frac{L_\gamma}{4\Gamma^2}\,,
\\
\label{eq_constr_Ljmin2}
L_{\rm j,B,min} &=& \left(\frac{\mathcal{D}}{\Gamma}\right)^4\frac{g_{\rm ERC}\Gamma^6\zeta(r)L_{\rm d}}{3q}\left[\frac{ct_{\rm var,obs}}{r(1+z)}\right]^2\,.
\eea
The dependence of the magnetic jet power on $\Gamma$ is much steeper than for the radiative jet power.
Thus, we can derive approximate constraints on $\Gamma$ in two limits.
For $u_\gamma' \gg u_{\rm B}'$ we find
\be
\label{eq_constr_Ljgammamin}
\Gamma(L_{\rm j,\gamma,min}) = \left(\frac{\mathcal{D}}{\Gamma}\right)^{-2}\left(\frac{L_\gamma}{4L_{\rm j,\gamma,min}}\right)^{1/2}\,;
\ee
and for $u_\gamma' \ll u_{\rm B}'$ we find
\bea
\label{eq_constr_LjBmin}
\Gamma(r,L_{\rm j,B,min}) &=& \left(\frac{\mathcal{D}}{\Gamma}\right)^{-2/3}\left(\frac{3q L_{\rm j,B,min}}{g_{\rm ERC}\zeta(r)L_{\rm d}}\right)^{1/6}
\times\nonumber\\&&
\left[\frac{r(1+z)}{ct_{\rm var,obs}}\right]^{1/3}
\,.
\eea
In Section \ref{sec_cases}, we will investigate the values of $u_\gamma'/u_{\rm B}'$ and $L_{\rm j,min}$ for individual blazar flares. Again, we stress that contributions from cold/warm electrons and protons should be included to obtain total jet energies.

\begin{figure*}[ht]
\centering
\includegraphics[width=0.8\textwidth]{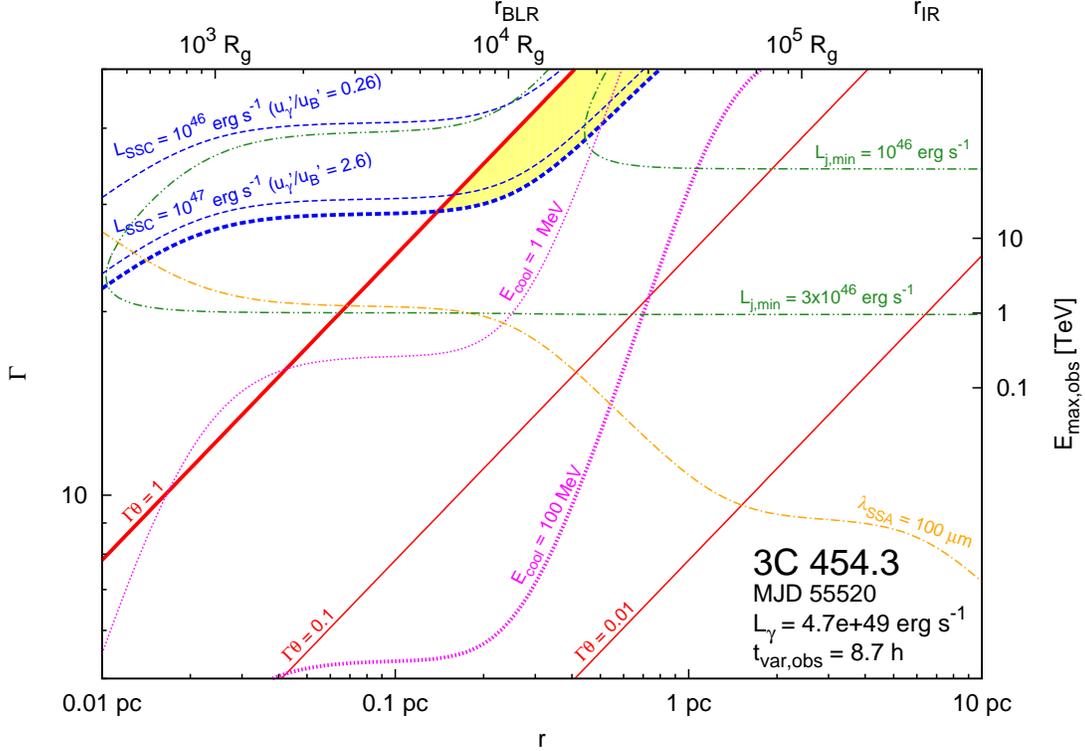}
\caption{Parameter space of distance scale $r$ and Lorentz factor $\Gamma$ of the emitting region responsible for the major gamma-ray flare of 3C~454.3 that peaked at MJD~55520.
Five classes of constraints are indicated:
the collimation constraint (\emph{solid red lines}; Eq. \ref{eq_constr_Gammatheta}),
the SSC constraint (\emph{dashed blue lines}; Eq. \ref{eq_constr_LSSC}),
the cooling constraint (\emph{dotted magenta lines}; Eq. \ref{eq_constr_Ecool}),
the synchrotron self-absorption constraint (\emph{dot-dashed orange lines}; Eq. \ref{eq_constr_nuabsobs}),
and the intrinsic gamma-ray opacity constraint (denoted by the maximum escaping photon energy labeled along the right-hand vertical axis; Eq. \ref{eq_constr_Emax}).
We also show predictions for the jet energetics:
the equipartition parameter ($u_{\rm\gamma}'/u_{\rm B}'$, shown together with the SSC constraint; Eq. \ref{eq_constr_ugammauB}),
and the minimum required jet power (\emph{double-dot-dashed green lines}; Eqs. \ref{eq_constr_Ljmin1}, \ref{eq_constr_Ljmin2}).
On the upper horizontal axis, we show the distance scale in terms of the gravitational radius of the supermassive black hole,
and the characteristic radii for main external radiation components (BLR and IR).
\emph{Yellow-shaded area} marks the parameter space allowed by the conditions $\Gamma\theta < 1$, $L_{\rm SSC} < L_{\rm X}$, and $E_{\rm cool,obs} < 100\;{\rm MeV}$.}
\label{fig_3c454.3_55520}
\end{figure*}

\section{Case studies}
\label{sec_cases}

In this section, we apply the constraints derived in Section \ref{sec_constr} to several well-studied cases of powerful gamma-ray flares in blazars with excellent multiwavelength coverage. We would like to emphasize the value of having extensive simultaneous spectral coverage of these sources, however, each case is different and the data quality is not uniform enough to warrant a broader study.

\subsection{3C~454.3 at MJD 55520}
\label{sec_3c454.3_55520}

3C~454.3 ($z = 0.859$, $d_{\rm L} \simeq 5.49\;{\rm Gpc}$) provided us with the most spectacular gamma-ray flares in the \emph{Fermi} era \citep{2013MNRAS.430.1324N}. On MJD~55520 (2010~Nov~20) it produced a flare of apparent peak bolometric ($E > 100\;{\rm MeV}$) luminosity of $L_{\rm\gamma,bol} \simeq 2.1\times 10^{50}\;{\rm erg\,s^{-1}}$ \citep{2011ApJ...733L..26A}.
We convert the bolometric peak luminosity $L_{\rm\gamma,bol}$ into the peak $\nu L_\nu$ luminosity $L_\gamma$, using a bolometric correction factor $g_{\rm\gamma,bol} = L_{\rm\gamma,bol}/L_\gamma \sim 4.5$ calculated from the best-fit spectral model (power-law with exponential cut-off), resulting in $L_\gamma \simeq 4.7\times 10^{49}\;{\rm erg\,s^{-1}}$.
The flare temporal template fitted by \cite{2011ApJ...733L..26A} has a flux doubling time scale of $t_{\rm var,obs} \simeq 8.7\;{\rm h} \simeq 3.13\times 10^4\;{\rm s}$.
\cite{2011ApJ...736L..38V} showed that this gamma-ray flare was accompanied by simultaneous outbursts, of amplitude smaller by factor $\sim 3$, in soft X-ray, optical and millimeter bands. They compiled an SED from which we can estimate the simultaneous luminosity ratios $q = L_{\rm\gamma}/L_{\rm syn} \simeq 30$, $L_{\rm syn}/L_{\rm X} \simeq 10$. These ratios are used to derive the simultaneous soft X-ray luminosity $L_{\rm X} \simeq 1.6\times 10^{47}\;{\rm erg\,s^{-1}}$.
We can also estimate the spectral index of the X-ray part of the spectrum as $\alpha \simeq 0.65$.
The bolometric accretion disk luminosity is taken as $L_{\rm d} \simeq 6.75\times 10^{46}\;{\rm erg\,s^{-1}}$ \citep{2011MNRAS.410..368B}, from which we find the characteristic radii of external radiation components $r_{\rm BLR} \simeq 0.26\;{\rm pc}$ and $r_{\rm IR} \simeq 6.5\;{\rm pc}$.
The black hole mass of 3C~454.3 is uncertain; here we adopt the value of $M_{\rm BH} \sim 5\times 10^8\,M_\sun$ after \cite{2011MNRAS.410..368B}.

In Figure \ref{fig_3c454.3_55520}, we plot the constraints on $r$ and $\Gamma$ corresponding to fixed values of $\Gamma\theta$, $L_{\rm SSC}$, $E_{\rm cool,obs}$, $\lambda_{\rm SSA,obs}$ and $E_{\rm max,obs}$, as well as the energetics parameters $u_\gamma'/u_{\rm B}'$ and $L_{\rm j,min}$.
We assumed here that $\xi_{\rm BLR} \simeq \xi_{\rm IR} \simeq 0.1$.
The yellow-shaded area is defined by the following 3 conditions: $\Gamma\theta < 1$, $L_{\rm SSC} < L_{\rm X}$, and $E_{\rm cool,obs} < 100\;{\rm MeV}$.
The intersection of the first two of these constraints gives the \emph{marginal solution} --- the minimum Lorentz factor $\Gamma_{\rm min} \simeq 30$ and the minimum distance scale $r_{\rm min} \simeq 0.16\;{\rm pc}$.
For $(r_{\rm min},\Gamma_{\rm min})$, other constraints yield the following predictions: $\lambda_{\rm SSA,obs} \simeq 125\;{\rm\mu m}$, $E_{\rm max,obs} \gtrsim 10\;{\rm TeV}$, $u_\gamma'/u_{\rm B}' \simeq 3.3$, and $L_{\rm j,min} \simeq 1.7\times 10^{46}\;{\rm erg\,s^{-1}} \simeq 0.25\;L_{\rm d}$.
On the other hand, in the IR region ($r \sim r_{\rm IR}$), the SSC constraint is much stronger and hence there are no solutions with $\Gamma < 50$.
Therefore, in this case the dissipation region is clearly constrained to be located not far from $r_{\rm BLR}$.
The minimum required jet power is one order of magnitude higher than the kinetic jet power estimated by \cite{2011ApJ...740...98M}. 

VLBI measurements of the jet of 3C~454.3 yield $\Gamma_{\rm j} \simeq 20$, $\mathcal{D} \simeq 33$ \citep{2009A&A...494..527H}, and $\Gamma_{\rm j}\theta_{\rm j} \simeq 0.3$ \citep{2009A&A...507L..33P}.
Adopting $\mathcal{D}/\Gamma_{\rm j} \simeq 1.67$ would shift the marginal solution to $r_{\rm min} \simeq 0.09\;{\rm pc}$ and $\Gamma_{\rm min} \simeq 18$.
The VLBI-derived solution of $r \simeq 0.34\;{\rm pc}$ and $\Gamma \simeq 20$ would be consistent with our $E_{\rm cool,obs}$ constraint, and marginally consistent with our $L_{\rm SSC}$ constraint.
On the other hand, for $\mathcal{D}/\Gamma = 1$, the SSC constraint also implies that jet collimation parameter is $\Gamma\theta > 0.5$.

\cite{2011ApJ...733L..26A} estimated the minimum Doppler factor of the emitting region responsible for this flare as $\mathcal{D}_{\rm min} \simeq 16$, using the gamma-ray opacity constraint for the maximum observed photon energy of $E_{\rm max,obs} = 31\;{\rm GeV}$.
Our opacity constraint for the same $E_{\rm max,obs}$ yields $\Gamma_{\rm min} = \mathcal{D}_{\rm min} \simeq 13$.
The main reason for this discrepancy is that we use the $3.6$ factor in Eq. (\ref{eq_Esoftobs}), which is neglected in numerous studies. 
We would like to point out that the SSC constraint is stronger than the opacity constraint \citep[see][]{2010ApJ...721.1383A}.
We also note that our minimum distance scale is compatible with the estimate of $r_{\rm min} \simeq 0.14\;{\rm pc}$ obtained by calculating gamma-ray opacity due to the broad-line photons \citep{2011ApJ...733L..26A}.

The synchrotron self-absorption break is predicted to fall in the far-IR range, both at the BLR and IR distance scales.
3C~454.3 was observed by Herschel PACS and SPIRE instruments during and after the peak of this gamma-ray flare \citep{2012ApJ...758...72W}.
While the period of the highest gamma-ray state was sparsely covered in the far-IR band, a very good correlation between the $160\;{\rm\mu m}$ data and the \emph{Fermi}/LAT gamma rays was found.
Such a correlation implies that the gamma-ray producing region is transparent to synchrotron self-absorption, i.e., that $\lambda_{\rm SSA,obs} \gtrsim 160\;{\rm\mu m}$.
Such a condition can be easily satisfied, together with our collimation and SSC constraints, even at BLR distance scales.
However, \cite{2012ApJ...758...72W} also showed that $1.3\,{\rm mm}$ data from SMA, of much better sampling rate, correlate well with the gamma rays.
This is very difficult to explain in a one-zone model --- the $1.3\;{\rm mm}$ synchrotron self-absorption line satisfies all three constraints only at $r \simeq 27\;{\rm pc}$ and $\Gamma \simeq 400$.
The most reasonable way to accommodate this observation is to consider a different variability time scale of the $1.3\;{\rm mm}$ emission.
Indeed, data presented in Fig. 10 of \cite{2012ApJ...758...72W} indicate that the flux-doubling time scale corresponding to the fastest observed increase of the $1.3\;{\rm mm}$ flux is $t_{\rm var,mm} \simeq 7.5\;{\rm d}$.
Adopting this variability time scale, the $1.3\;{\rm mm}$ photosphere can be located already at $r_{\rm mm} \simeq 5\;{\rm pc}$ and $\Gamma = 38$, which are much more reasonable parameters.
Therefore, we need to consider an extended, possibly structured emitting region for the gamma rays observed during this event, with the rapidly flaring component produced at sub-pc scales, and a more slowly varying component correlated with the $1.3\;{\rm mm}$ emission at supra-pc scales.
This scenario is similar to the one proposed for PKS 1510-089 by \cite{2012ApJ...760...69N}.

\begin{figure*}[ht]
\centering
\includegraphics[width=0.8\textwidth]{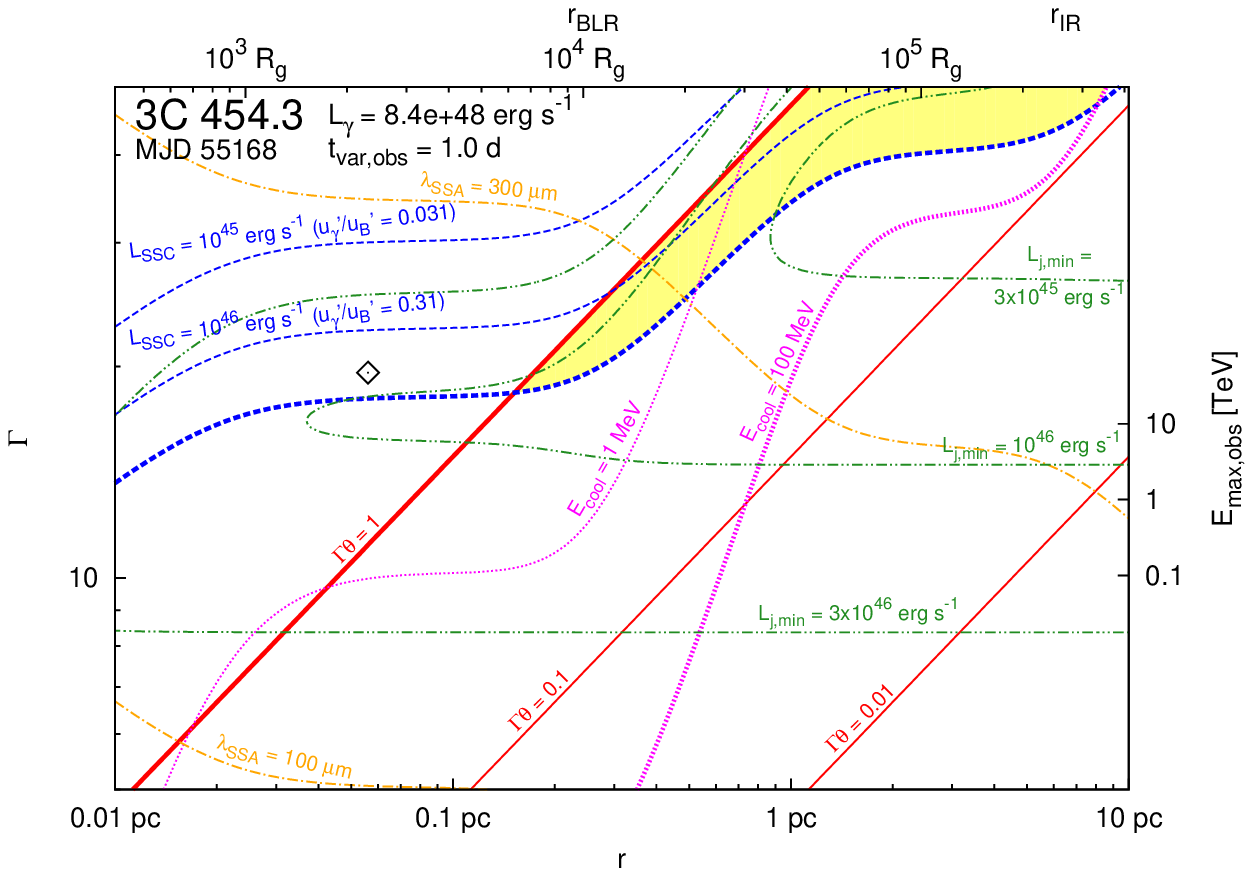}
\caption{Parameter space of $r$ and $\Gamma$ for the major flare of 3C~454.3 that peaked at MJD~55168. See Fig. \ref{fig_3c454.3_55520} for detailed description. The \emph{diamond} indicates the solution obtained by \cite{2011MNRAS.410..368B}.}
\label{fig_3c454.3_55168}
\end{figure*}

\subsection{3C~454.3 at MJD 55168}
\label{sec_3c454.3_55168}

A previous flare of 3C~454.3, peaking at MJD~55168 (2009~Dec~3), also attracted considerable interest \citep[\eg,][]{2010ApJ...716L.170P, 2010ApJ...721.1383A, 2011MNRAS.410..368B}.
The apparent peak bolometric gamma-ray luminosity was $L_{\rm\gamma,bol} \simeq 3.8\times 10^{49}\;{\rm erg\,s^{-1}}$ \citep{2010ApJ...721.1383A}, which corresponds to the $\nu L_\nu$ luminosity $L_\gamma = L_{\rm\gamma,bol}/g_{\rm\gamma,bol} \simeq 8.4\times 10^{48}\;{\rm erg\,s^{-1}}$.
The variability time scale was estimated at $t_{\rm var,obs} \simeq 1\;{\rm d}$, although episodes were observed with a flux-doubling time scale as short as $\simeq 2.3\;{\rm h}$.
From the SED compiled by \cite{2011MNRAS.410..368B}, we deduce $q = L_\gamma/L_{\rm syn} \simeq 14$, $L_{\rm syn}/L_{\rm X} \simeq 10$, and $\alpha \simeq 0.55$.
We use the same values of $\xi_{\rm BLR}$, $\xi_{\rm IR}$, $L_{\rm d}$, and $M_{\rm BH}$ as for the MJD~55520 flare.

Our constraints for this event are shown in Figure \ref{fig_3c454.3_55168}.
We find the marginal solution at $r_{\rm min} \simeq 0.17\;{\rm pc}$ and $\Gamma_{\rm min} \simeq 19$.
This solution corresponds to $\lambda_{\rm SSA,obs} \simeq 215\;{\rm\mu m}$, $u_\gamma'/u_{\rm B}' \simeq 1.6$, and $L_{\rm j,min} \simeq 10^{46}\;{\rm erg\,s^{-1}} \sim 0.14 L_{\rm d}$.
While solutions within $r_{\rm BLR}$ are allowed, the maximum observed photon energy is $E_{\rm max,obs} \simeq 21\;{\rm GeV}$ \citep{2010ApJ...721.1383A}, so the effect of BLR absorption is expected to be lower than in the case of the MJD~55520 flare (Section \ref{sec_3c454.3_55520}).

\cite{2011MNRAS.410..368B} modeled the SEDs of 3C~454.3 for several epochs close to MJD~55168, probing different luminosity levels.
They noted that the gamma-ray luminosity scales with the X-ray and UV luminosities roughly like $L_\gamma \propto L_X^2 \propto L_{\rm UV}^2$.
Therefore, they proposed that the location of the gamma-ray emitting region shifts outwards with increasing gamma-ray luminosity.
For the highest state at MJD~55168, they suggested a distance scale of $r \simeq 0.06\;{\rm pc}$ at $\Gamma \simeq 20$ (see Figure \ref{fig_3c454.3_55168}).
It is critical to note at this point that they adopted a variability time scale of $t_{\rm var,obs} \simeq 6\;{\rm h}$, and a Doppler-to-Lorentz factor ratio of $\mathcal{D}/\Gamma \simeq 1.45$.
We have checked that for such parameters our constraints are marginally consistent with their result;
our model predicts $u_\gamma'/u_{\rm B}' \simeq 0.84$, $\lambda_{\rm SSA} \simeq 118\;{\rm\mu m}$, and $L_{\rm j,min} \simeq 2.7\times 10^{45}\;{\rm erg\,s^{-1}}$.

\begin{figure*}[ht]
\centering
\includegraphics[width=0.8\textwidth]{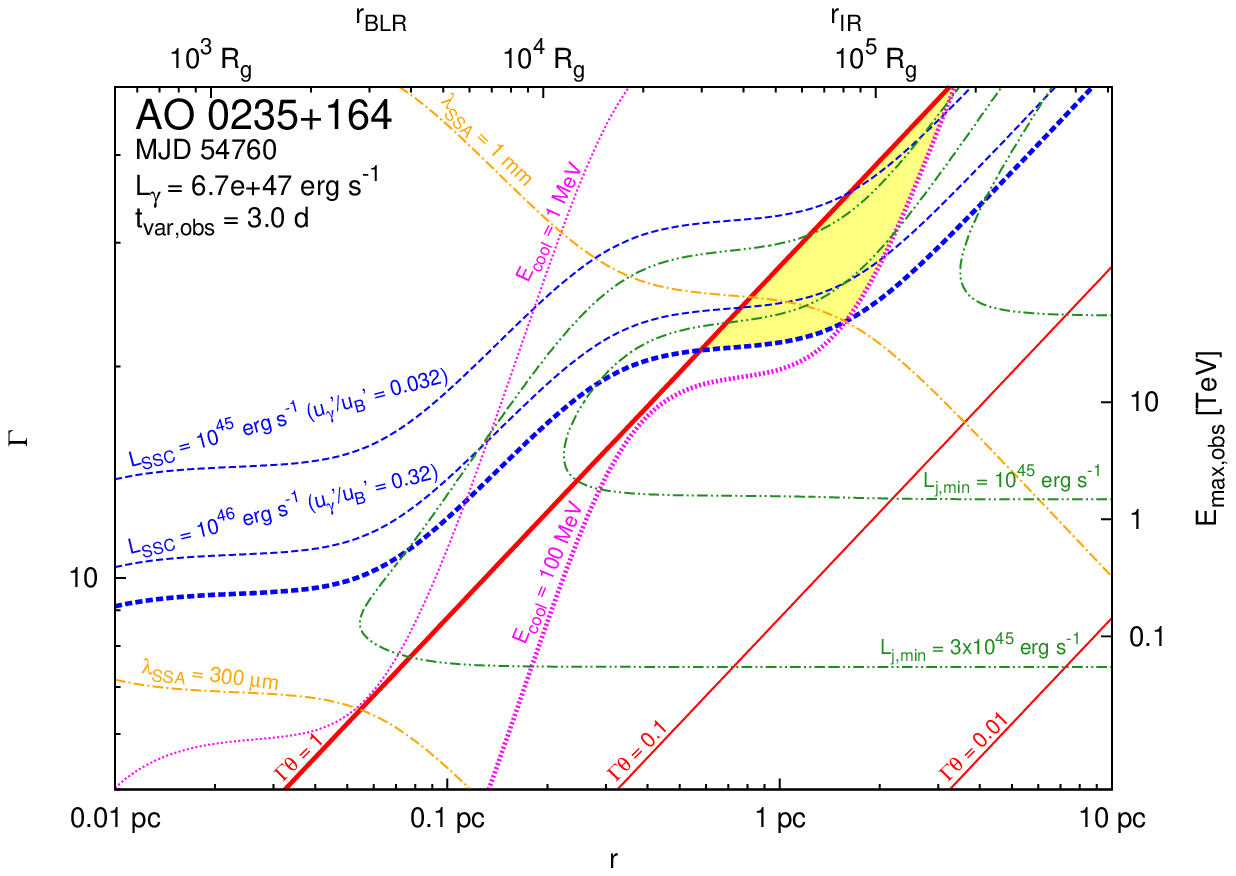}
\caption{Parameter space of $r$ and $\Gamma$ for the major flare of AO~0235+164 that peaked at MJD~54760. See Fig. \ref{fig_3c454.3_55520} for detailed description.}
\label{fig_ao0235}
\end{figure*}

\subsection{AO~0235+164 at MJD 54760}
\label{sec_ao0235}

AO~0235+164 ($z = 0.94$, $d_{\rm L} \simeq 6.14\;{\rm Gpc}$) is an LBL-type blazar, which was active in 2008--2009.
The highest gamma-ray state, achieved between MJD~54700 and MJD~54780, was analyzed in detail by \cite{2012ApJ...751..159A}.
They estimated the observed gamma-ray luminosity as $L_\gamma \simeq 6.7\times 10^{47}\;{\rm erg\,s^{-1}}$;
the observed variability time scale $t_{\rm var,obs} \simeq 3\;{\rm d} = 2.6\times 10^5\;{\rm s}$;
the Compton dominance $q = L_\gamma/L_{\rm syn} \simeq 4$;
the synchrotron to X-ray luminosity ratio $L_{\rm syn}/L_{\rm X} \simeq 6$;
the accretion disk luminosity $L_{\rm d} = 4\times 10^{45}\;{\rm erg\,s^{-1}}$;
and the characteristic radii of external radiation components $r_{\rm BLR} \simeq 0.06\;{\rm pc}$ and $r_{\rm IR} \simeq 1.6\;{\rm pc}$.
For the black hole mass, they adopted $M_{\rm BH} \sim 4\times 10^8\,M_\sun$.
The X-ray spectral index is very uncertain, very soft X-ray spectra were observed by \emph{Swift}/XRT during the gamma-ray activity. Here we adopt $\alpha \simeq 1$.

In Figure \ref{fig_ao0235}, we plot the constraints on the location of the gamma-ray flare, adopting $\xi_{\rm BLR} = \xi_{\rm IR} = 0.1$.
The marginal solution is located at $r_{\rm min} \simeq 0.65\;{\rm pc}$ and $\Gamma_{\rm min} \simeq 22$.
The predictions for this solution are $\lambda_{\rm SSA,obs} \simeq 920\;{\rm\mu m}$, $u_\gamma'/u_{\rm B}' \simeq 0.7$, and $L_{\rm j,min} \simeq 8.5\times 10^{44}\;{\rm erg\,s^{-1}} \simeq 0.2 L_{\rm d}$.
The gamma-ray emitting region is certainly located outside the BLR, in the region where external radiation is dominated by the dusty torus emission.
The jet is predicted to be at least moderately magnetized at $r \sim 3\times 10^4\;R_{\rm g}$.
The required minimum jet power is higher by factor $\simeq 4$ than the estimate of \cite{2011ApJ...740...98M}.

VLBI measurements of the jet of AO~0235+164 imply that $\mathcal{D}/\Gamma_{\rm j} \simeq 1.98$ and $\Gamma_{\rm j}\theta_{\rm j} \simeq 0.04$ \citep{2009A&A...494..527H,2009A&A...507L..33P}.
This rather extreme solution of a very narrow and perfectly aligned jet is inconsistent with both the $L_{\rm SSC}$ and $E_{\rm cool,obs}$ constraints.
For $\mathcal{D}/\Gamma = 1$, the combination of $L_{\rm SSC}$ and $E_{\rm cool,obs}$ constraints implies that $\Gamma\theta > 0.4$.

\cite{2011ApJ...735L..10A} presented a detailed discussion of the same event, and they argued that this flare was produced at the distance scale of $\sim 12\;{\rm pc}$, based on the VLBI imaging and cross-correlation between the gamma rays and the mm data.
\cite{2012ApJ...751..159A} used a simple variability time scale argument to show that locating the emitting region at $12\;{\rm pc}$ would require a very high jet Lorentz factor $\Gamma \simeq 50$.
Here, we find that the SSC constraint leads to a similar limit on $\Gamma$ already at $r \simeq 9\;{\rm pc}$.
Moreover, the cooling constraint is even stronger at distances larger than $\simeq r_{\rm IR}$, implying that energetic electrons injected at the distance of $12\;{\rm pc}$ have no chance to cool down efficiently.
On the other hand, we show that if the emitting region is located at $r_{\rm IR}$ and has a moderate Lorentz factor of $\Gamma \simeq 24$, it will be transparent to wavelengths shorter than $\simeq 1\;{\rm mm}$.
\cite{2011ApJ...735L..10A} calculated the discrete correlation function (DCF) between the gamma rays and the $1\;{\rm mm}$ light curve, showing multiple peaks in the range of delays between $0$ and $-50$ days (the latter meaning that the gamma rays lead the mm signals).
Our result is thus not in conflict with the gamma -- $1\;{\rm mm}$ DCF.
However, our model does not allow for the possibility that the emitting region producing $3$-day long gamma-ray flares is transparent at $7\;{\rm mm}$, which is the wavelength of VLBA observations reported by \cite{2011ApJ...735L..10A}.
In our model, even for $\Gamma = 100$ the $7\;{\rm mm}$ photosphere would fall at a very large distance of $\simeq 90\;{\rm pc}$.
Just like in the case of 3C~454.3 (see Section \ref{sec_3c454.3_55520}), the solution to this apparent paradox is that the variability time scale of the $7\;{\rm mm}$ radiation has to be much longer than 3 days.
Indeed, the $7\;{\rm mm}$ light curves presented in \cite{2011ApJ...735L..10A} indicate variability time scale of the order of $\simeq 80$ days.
When we used this time scale to calculate the collimation ($\Gamma\theta$) and the synchrotron self-absorption ($\lambda_{\rm SSA,obs}$) constraints, we obtained the following solution: the $\Gamma\theta = 1$ line crosses the $7\;{\rm mm}$ photosphere at $r_{\rm 7mm} \simeq 6.7\;{\rm pc}$ and $\Gamma_{\rm j,7mm} \simeq 14$.
This is consistent with the detection around this epoch of a superluminal radio element of apparent velocity $\beta_{\rm app} \sim 13$ \citep{2011ApJ...735L..10A}.

The close observed correspondence between the gamma-ray flares and the activity at the $7\;{\rm mm}$ wavelength does not necessarily indicate that the gamma rays should be produced co-spatially with the $7\;{\rm mm}$ core.
In Appendix \ref{sec_disc_gamma_mm}, we present a simple light travel time argument according to which the gamma rays could still be produced at the distance of $\sim 1\;{\rm pc}$.

Our results indicate that the $12\;{\rm pc}$ scenario cannot be constrained by energetic requirements, as the required minimum jet power is only $L_{\rm j,min} \sim 3\times 10^{44}\;{\rm erg\,s^{-1}}$ in this case.
However, even a moderate jet magnetization implied by the SSC constraint puts into question the efficiency of the reconfinement/conical shock that is proposed by \cite{2011ApJ...735L..10A} as the physical mechanism behind the $7\;{\rm mm}$ core.

\begin{figure*}[ht]
\centering
\includegraphics[width=0.8\textwidth]{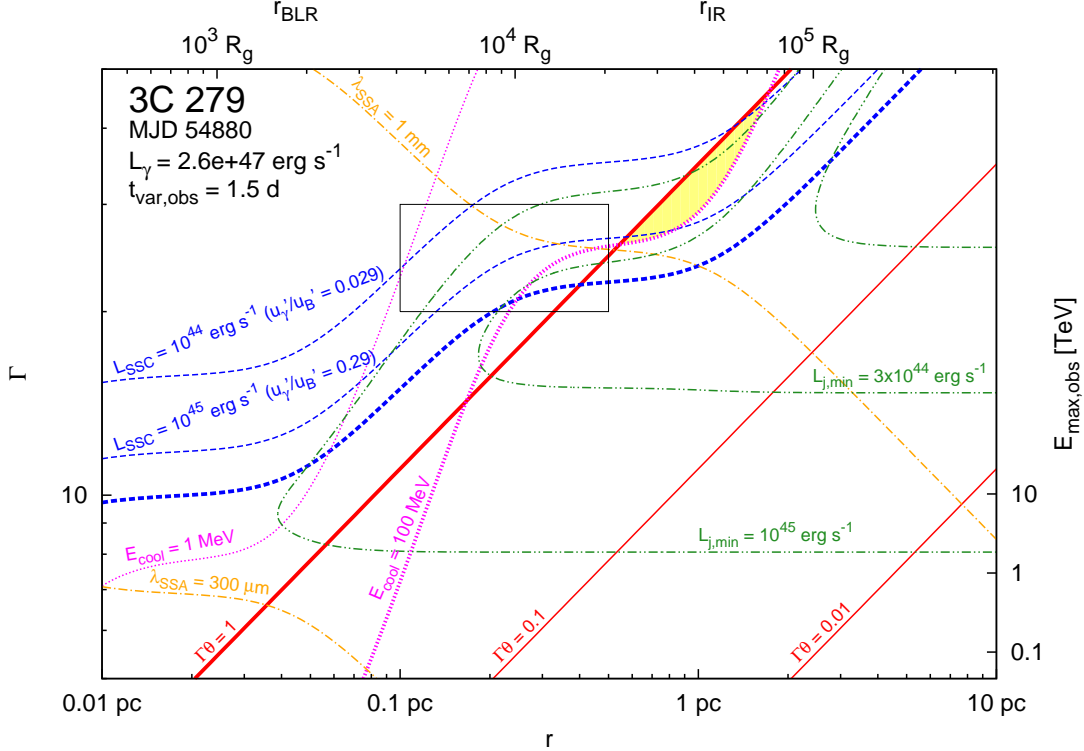}
\caption{Parameter space of $r$ and $\Gamma$ for the major flare of 3C~279 that peaked at MJD~54880. See Fig. \ref{fig_3c454.3_55520} for detailed description. The \emph{black box} indicates roughly the parameter space region constrained by \cite{2014ApJ...782...82D}.}
\label{fig_3c279}
\end{figure*}

\subsection{3C~279 at MJD~54880}
\label{sec_3c279}

3C~279 ($z=0.536$, $d_{\rm L} \simeq 3.07\;{\rm Gpc}$) produced a gamma-ray flare peaking at MJD~54880 that was extensively studied in \cite{2010Natur.463..919A} and \cite{2012ApJ...754..114H}.
The gamma-ray flux doubling time scale can be estimated as $t_{\rm var,obs}\simeq 1.5\;{\rm d}$, and the half-peak gamma-ray luminosity is $L_\gamma \simeq 2.6\times 10^{47}\;{\rm erg\,s^{-1}}$.
Following \cite{2012ApJ...754..114H}, we adopt $L_{\rm d} \simeq 2\times 10^{45}\;{\rm erg\,s^{-1}}$, $q \simeq 7.5$, $L_{\rm syn}/L_{\rm X} \simeq 9.2$, $M_{\rm BH} \simeq 5\times 10^8\,M_\sun$, and $\alpha \simeq 0.7$.
This implies that $r_{\rm BLR} \simeq 0.045\;{\rm pc}$ and $r_{\rm IR} \simeq 1.1\;{\rm pc}$.

In Figure \ref{fig_3c279}, we plot the constraints on $r$ and $\Gamma$ for this flare.
The marginal solution is $r_{\rm min} \simeq 0.62\;{\rm pc}$ and $\Gamma_{\rm min} \simeq 27$, which locates the gamma-ray emission firmly outside the BLR, and close to $r_{\rm IR}$.
The predictions for this solution are $\lambda_{\rm SSA,obs} \sim 1.03\;{\rm mm}$, $u_\gamma'/u_{\rm B}' \sim 0.3$, $L_{\rm j,min} \sim 4\times 10^{44}\;{\rm erg\,s^{-1}} \sim 0.2 L_{\rm d}$.
The required jet power is roughly half of the estimate of \cite{2011ApJ...740...98M}.

The MOJAVE jet kinematics solution yields $\mathcal{D} \simeq 24$, $\Gamma_{\rm j} \simeq 21$ \citep{2009A&A...494..527H}, and $\Gamma_{\rm j}\theta_{\rm j} \simeq 0.22$ \citep{2009A&A...507L..33P}.
The implied Doppler-to-Lorentz factor ratio of $\mathcal{D}/\Gamma \simeq 1.15$ is fairly close to unity.
This solution is inconsistent with both the $L_{\rm SSC}$ and $E_{\rm cool,obs}$ constraints.
For $\mathcal{D}/\Gamma = 1$, the combination of the $L_{\rm SSC}$ and $E_{\rm cool,obs}$ constraints implies that $\Gamma\theta > 0.7$.

\cite{2012ApJ...754..114H} proposed two scenarios for the gamma-ray emission.
One of them emphasized the connection to a $20\;{\rm d}$-scale polarization event, which implicated the location at $1-4\;{\rm pc}$.
The other was based on mid-IR spectral structure detected by \emph{Spitzer}, which was interpreted as a synchrotron self-absorption turnover.
The latter implicated sub-pc scales ($r_{\rm BLR}$) for the main synchrotron/gamma-ray component, with an additional emitting region located at $\sim 4\;{\rm pc}$.
Our results show very clearly that location of the gamma-ray flare at $r_{\rm BLR}$ is not consistent with the variability time scale of days, rather it would require a variability time scale of several hours.
With the relatively moderate peak gamma-ray flux of 3C~279, such short time scales could not be probed with \emph{Fermi}/LAT.
Such time scales are essential in order to interpret the \emph{Spitzer} spectral feature in terms of synchrotron self-absorption.
On the other hand, the distance of $1\;{\rm pc}$ is fully consistent with all constraints, however, shifting the emitting region to the distance of $4\;{\rm pc}$ would violate the $E_{\rm cool,obs}$ constraint.

\cite{2014ApJ...782...82D} presented a detailed model of the radiation of blazars which was applied to the 3C~279 data from \cite{2012ApJ...754..114H}.
They concluded that this gamma-ray flare was produced at $r \sim 0.1 - 0.5\;{\rm pc}$ for $\Gamma \sim 20-30$.
This is still outside the BLR, but according to Figure \ref{fig_3c279} their parameter region extends well into the $\Gamma\theta > 1$ regime.
However, they assumed a very short variability time scale of $t_{\rm var,obs} \sim 10^4\;{\rm s} = 2.8\;{\rm h}$.
We have checked the consequences of adopting $t_{\rm var} = 10^4\;{\rm s}$ in our model.
For $20 \le \Gamma \le 30$, we found a range of possible locations $r \sim 0.025 - 0.11\;{\rm pc}$, which are closer to the black hole than the solutions of \cite{2014ApJ...782...82D}.
In that work, the location of the gamma-ray emitting region was constrained by calculating $u_{\rm BLR}$ from SED modeling, and comparing it with the level $u_{\rm BLR,0}$ expected for $r < r_{\rm BLR}$.
By noting that $u_{\rm BLR} < u_{\rm BLR,0}$, they concluded that $r > r_{\rm BLR}$.
However, it is difficult to provide a precise estimate of $r$ in this way, because it depends on the uncertain shape of the $u_{\rm BLR}'(r)$ function for $r > r_{\rm BLR}$.
Because these authors allowed for higher values of the accretion disk luminosity, up to $L_{\rm d} = 10^{46}\;{\rm erg\,s^{-1}}$, they also have higher values of $r_{\rm BLR} \propto L_{\rm d}^{1/2} \lesssim 0.1\;{\rm pc}$.
Taking these differences into account, the discrepancy between their and our results does not appear to be significant.

\begin{figure*}[ht]
\centering
\includegraphics[width=0.8\textwidth]{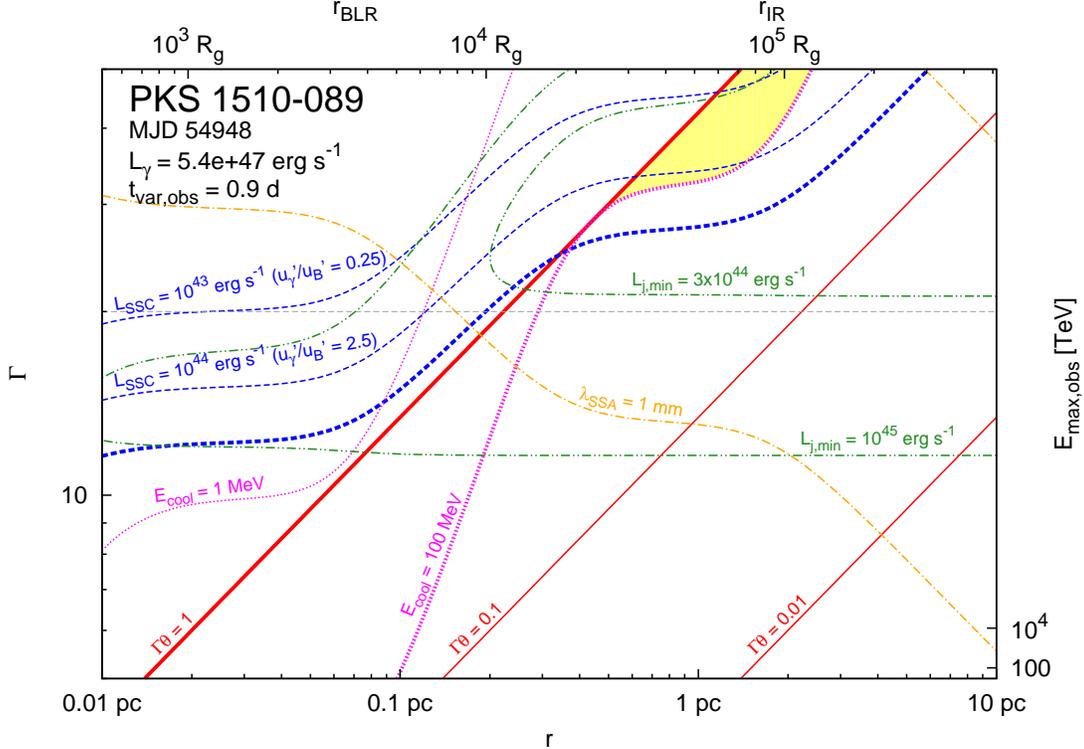}
\caption{Parameter space of $r$ and $\Gamma$ for the major flare of PKS~1510-089 that peaked at MJD~54948. See Fig. \ref{fig_3c454.3_55520} for detailed description.}
\label{fig_pks1510}
\end{figure*}

\subsection{PKS~1510-089 at MJD~54948}
\label{sec_pks1510}

PKS~1510-089 ($z = 0.36$, $d_{\rm L} \simeq 1.92\;{\rm Gpc}$), the second most active blazar of the \emph{Fermi} era \citep{2013MNRAS.430.1324N}, has been monitored extensively in the X-ray, optical/NIR, and radio/mm bands.
In early 2009, it produced a series of gamma-ray flares, peaking at MJD~54917 (2009 Mar 27), MJD~54948 (2009 Apr 27), and MJD~54962 (2009 May 11) \citep{2010ApJ...721.1425A,2011A&A...529A.145D}.
The first and the last of them were accompanied by sharp optical/UV flares, but none of them had a clear X-ray counterpart.
A cross-correlation analysis indicates that the optical signal could be delayed with respect to the gamma-ray signal by $\simeq 13\;{\rm d}$, in which case the major optical flare peaking at MJD~54961 would be associated with the second gamma-ray event at MJD~54948.
However, in our work we are primarily concerned with the gamma-ray emitting regions as they are when they produce a gamma-ray flare, and thus we use strictly simultaneous multiwavelength data.
Therefore, we will focus on the case of MJD~54948, ignoring the optical flare that follows it.
As usual, there is some ambiguity about establishing the flare parameters, and for this purpose we carefully examine the results of \cite{2010ApJ...721.1425A}, and compare them with our own analysis.
We adopt the $\nu L_\nu$ gamma-ray luminosity of $L_\gamma \simeq 5.4\times 10^{47}\;{\rm erg\,s^{-1}}$, the gamma-ray variability time scale of $t_{\rm var,obs} \simeq 0.9\;{\rm d}$ \citep{2013MNRAS.430.1324N}, the accretion disk luminosity of $L_{\rm d} \simeq 5\times 10^{45}\;{\rm erg\,s^{-1}}$ \citep{2012ApJ...760...69N}, the Compton dominance parameter of $L_\gamma/L_{\rm syn} \simeq 100$, the X-ray luminosity of $L_X \simeq 5\times 10^{44}\;{\rm erg\,s^{-1}}$, the X-ray spectral index of $\alpha \simeq 0.3$, the black hole mass of $M_{\rm BH} \simeq 4\times 10^8\;M_\sun$, the covering factors of $\xi_{\rm BLR} = \xi_{\rm IR} \simeq 0.1$, and the external radiation fields radii $r_{\rm BLR} \simeq 0.07\;{\rm pc}$ and $r_{\rm IR} \simeq 1.8\;{\rm pc}$.

Our constraints for the MJD~54948 flare of PKS~1510-089 are presented in Figure \ref{fig_pks1510}.
The SSC constraint is particularly strong in this case, since $L_\gamma/L_X \simeq 1000$.
The marginal solution is $r_{\rm min} \simeq 0.37\;{\rm pc}$ at $\Gamma_{\rm min} \simeq 26$, which is well outside the BLR.
The predictions for this solution are $\lambda_{\rm SSA,obs} \simeq 1.4\;{\rm mm}$, $u_\gamma'/u_{\rm B}' \simeq 12$, and $L_{\rm j,min} \simeq 2.2 \times 10^{44}\;{\rm erg\,s^{-1}} \sim 0.045\;L_{\rm d}$, which is slightly lower than the total jet power estimate by \cite{2011ApJ...740...98M}.
Therefore, we suggest that the jet of PKS~1510-089 is only weakly magnetized.

\cite{2010ApJ...721.1425A} argued that this gamma-ray flare was produced within the BLR, as they found that the gamma-ray and optical luminosities are related roughly like $L_\gamma \propto L_{\rm opt}^{1/2}$, which favors the ERC(BLR) mechanism of gamma-ray production over ERC(IR).
Their SED models were calculated for $\Gamma \simeq 15$, and their SSC components peak significantly below $L_X$.
This would be in strong disagreement with our results, if not for two crucial assumptions: they adopted $\mathcal{D}/\Gamma \simeq 1.4$ and $t_{\rm var} \simeq 0.25\;{\rm d}$.
When these parameters are used in our model, we obtain $r_{\rm min} \simeq 0.035\;{\rm pc}$ at $\Gamma_{\rm min} \simeq 12$, which is consistent with their result.
We note that VLBI observations indicate that $\mathcal{D}/\Gamma \simeq 0.8$ \citep{2009A&A...494..527H}, so our choice of $\mathcal{D}/\Gamma = 1$ seems to be more conservative.
\cite{2010ApJ...721.1425A} used the intrinsic gamma-ray opacity constraint to derive a limit on the Doppler factor $\mathcal{D} \gtrsim 8$, which we find very conservative, and certainly weaker than the SSC constraint.
They also estimated the jet power, and for this particular flare they obtained $L_{\rm j} \simeq 4.8\times 10^{45}\;{\rm erg\,s^{-1}}$, about $60\%$ of which is in the magnetic form, and only $\sim 8\%$ in the radiative form.
This indicates that in their model $u_\gamma'/u_{\rm B}' \simeq 0.13$, which is consistent with their low $L_{\rm SSC}$, but this solution is likely to require $\Gamma\theta > 1$.
The energetic requirements discussed by \cite{2010ApJ...721.1425A} can be significantly relaxed by bringing their model closer to equipartition.

\cite{2010ApJ...710L.126M} presented an independent analysis of the activity of PKS~1510-089 in early 2009, including more detailed VLBI analysis and optical polarization data.
The VLBI observations at $43\;{\rm GHz}$ revealed a superluminal knot of apparent velocity $22c$, which was projected to pass the stationary core at MJD $\sim 54959$, simultaneous with the major optical flare.
This optical flare was accompanied by a sharp increase of the optical polarization degree, up to $\sim 37\%$, and apparently preceded by a gradual ($\sim 50\;{\rm d}$ time scale) rotation of the optical polarization angle by $\sim 720^\circ$.
They interpreted the gamma-ray activity of PKS~1510-089 as directly related to the emergence of the superluminal radio/mm feature, with optical polarization rotation indicating either stochastic or helical structure of the jet.
This interpretation implies a $\sim 10-20\;{\rm pc}$ distance scale for the gamma-ray flares, at which the ERC mechanism based on IR photons is inefficient.
Instead, it was proposed that the gamma rays are produced by Comptonization of synchrotron radiation produced in slower outer jet layers \citep[spine-sheath models,][]{2005A&A...432..401G}.
In Appendix \ref{sec_disc_spine_sheath}, we show that in fact the spine-sheath model offers no advantage over the ERC model in explaining strongly beamed gamma-ray emission.

\begin{figure*}[ht]
\centering
\includegraphics[width=0.8\textwidth]{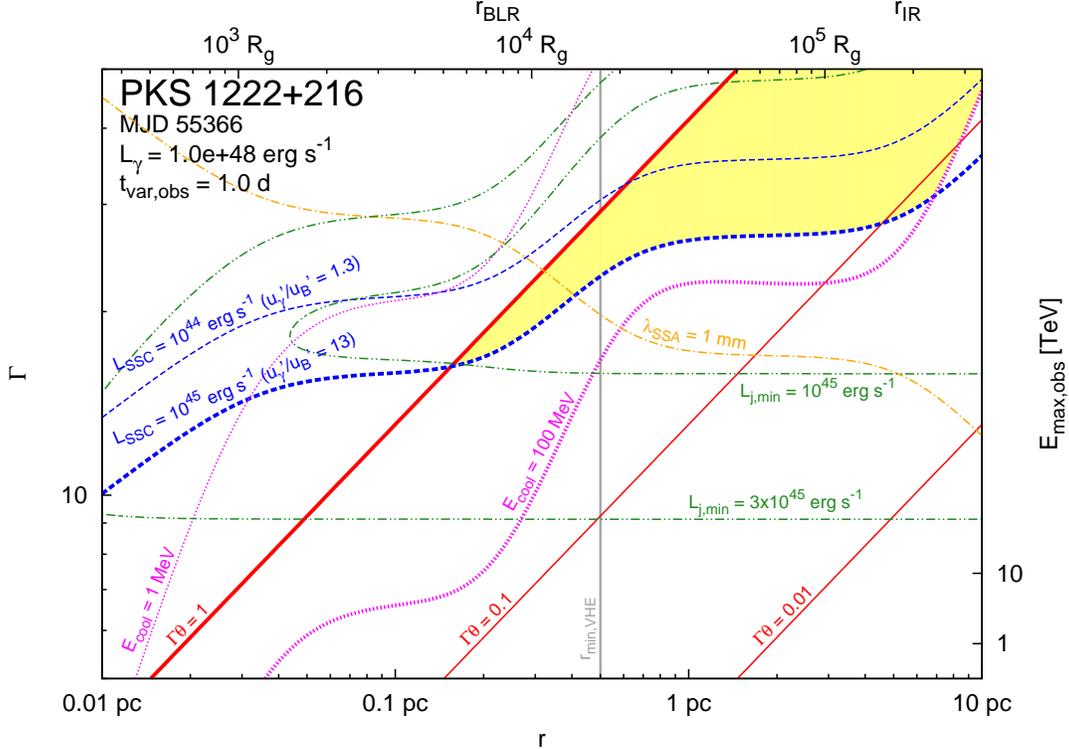}
\caption{Parameter space of $r$ and $\Gamma$ for the major flare of PKS~1222+216 that peaked at MJD~55366. See Fig. \ref{fig_3c454.3_55520} for detailed description. The \emph{vertical solid gray line} indicates the minimum distance for the production of VHE radiation observed by MAGIC.}
\label{fig_pks1222}
\end{figure*}

\cite{2012MNRAS.424..789C} performed time-dependent SED modeling of the March 2009 flare of PKS~1510-089, investigating three scenarios for the gamma-ray emission: ERC(BLR), ERC(IR), and SSC.
The ERC(BLR) scenario was demonstrated to require very low values of the covering factor, $\xi_{\rm BLR} \sim 0.01$.
The other two scenarios produce reasonable fits to the observed SEDs, each scenario having its own moderate problems.
The problem of localization of the gamma-ray emitting region was not directly addressed.
We note that since the ERC(BLR) model should be located at $r \lesssim r_{\rm BLR}$, it requires $\Gamma\theta \gg 1$, especially for the adopted variability time scale of $4\,{\rm d}$.
The SSC models are difficult to localize, because their parameters are independent of the external radiation fields.
However, in order to suppress the ERC component, they require a significantly lower Lorentz factor, $\Gamma \lesssim 10$, than the ERC models.
We discuss briefly the constraints on SSC models in Section \ref{sec_disc_limerc}.

During the active state in 2009, PKS~1510-089 was detected in the Very High Energy (VHE) gamma-ray band, up to $300\;{\rm GeV}$, by the H.E.S.S. observatory \citep{2013A&A...554A.107H}. Opacity constraints due to broad emission lines imply that the VHE emission must be produced outside the BLR \citep{2013arXiv1307.1779B}, which is fully consistent with our results for the GeV emission.

\subsection{PKS~1222+216 at MJD~55366}
\label{sec_pks1222}

PKS~1222+216 ($z = 0.432$, $d_{\rm L} \simeq 2.4\;{\rm Gpc}$) was in a very active gamma-ray state in 2010, producing major GeV flares peaking at MJD~55317 (2010 May 1) and MJD~55366 (2010 Jun 19) \citep{2011ApJ...733...19T}. Shortly before the latter event, MAGIC observatory detected VHE emission (up to $400\;{\rm GeV}$) of extremely short variability time scale, $\sim 9\;{\rm min}$ \citep{2011ApJ...730L...8A}, which proved to be very challenging to explain \citep{2011A&A...534A..86T,2012ApJ...755..147D,2012MNRAS.425.2519N,2012PhRvD..86h5036T,2013MNRAS.431..355G}. Arguably, the only certain result concerning this VHE event is that it should be produced at the distance scale beyond $r_{\rm min,VHE} \sim 0.5\;{\rm pc}$ in order to avoid the absorption of the VHE photons by the BLR radiation.
Here, we focus on the GeV flare peaking at MJD~55366, for which the variability time scale was estimated as $t_{\rm var,obs} \simeq 1\;{\rm d}$, and the gamma-ray luminosity as $L_\gamma \simeq 10^{48}\;{\rm erg\,s^{-1}}$ \citep{2011ApJ...733...19T}.
Following \cite{2011A&A...534A..86T}, we adopt $L_{\rm d} \simeq 5\times 10^{46}\;{\rm erg\,s^{-1}}$, $\xi_{\rm BLR} \simeq 0.02$, $\xi_{\rm IR} \simeq 0.2$, $q = L_\gamma/L_{\rm syn} \gtrsim 100$, $L_{\rm X} \simeq 10^{45}\;{\rm erg\,s^{-1}}$, $\alpha \simeq 0.6$, $r_{\rm BLR} \simeq 0.22\;{\rm pc}$, and $r_{\rm IR} \simeq 5.6\;{\rm pc}$.
There is significant uncertainty in the value of $q$, as the simultaneous \emph{Swift}/UVOT spectra are dominated by the thermal component.
The black hole mass was recently estimated as $M_{\rm BH} \simeq 6\times 10^8M_\sun$ \citep{2012MNRAS.424..393F}.

Our constraints for the GeV flare of PKS~1222+216 are presented in Figure \ref{fig_pks1222}.
The marginal solution is found at $r_{\rm min} \simeq 0.18\;{\rm pc}$ and $\Gamma_{\rm min} \simeq 17$.
This location is within the BLR, and significantly closer to the black hole than the minimum location of the VHE emission.
The predictions for this solution are: $\lambda_{\rm SSA,obs} \simeq 0.76\;{\rm mm}$, $u_\gamma'/u_{\rm B}' \simeq 11$, and $L_{\rm j,min} \simeq 9.5\times 10^{44}\;{\rm erg\,s^{-1}} \simeq 0.019\,L_{\rm d}$, which is slightly above the estimate by \cite{2011ApJ...740...98M}.
When we increase the Compton dominance parameter to $300$, we obtain $r_{\rm min} \simeq 0.13\;{\rm pc}$ and $\Gamma_{\rm min} \simeq 14$.
And when we use a shorter variability time scale of $\simeq 6\;{\rm h}$ \citep{2011arXiv1110.4471F}, we obtain $r_{\rm min} \simeq 0.08\;{\rm pc}$ and $\Gamma_{\rm min} \simeq 23$.

The VLBI kinematic solution is rather peculiar, with $\mathcal{D}/\Gamma_{\rm j} \simeq 0.11$ \citep{2009A&A...494..527H}, which would indicate that PKS~1222+216 is not a blazar.
When we decrease our Doppler-to-Lorentz factor ratio merely to $\mathcal{D}/\Gamma = 0.5$, a minimum Lorentz factor of $\Gamma_{\rm min} \simeq 52$ is required.
Therefore, adopting $\mathcal{D}/\Gamma \simeq 1$ seems to be the most reasonable option in this case.

\begin{figure*}[ht]
\centering
\includegraphics[width=0.8\textwidth]{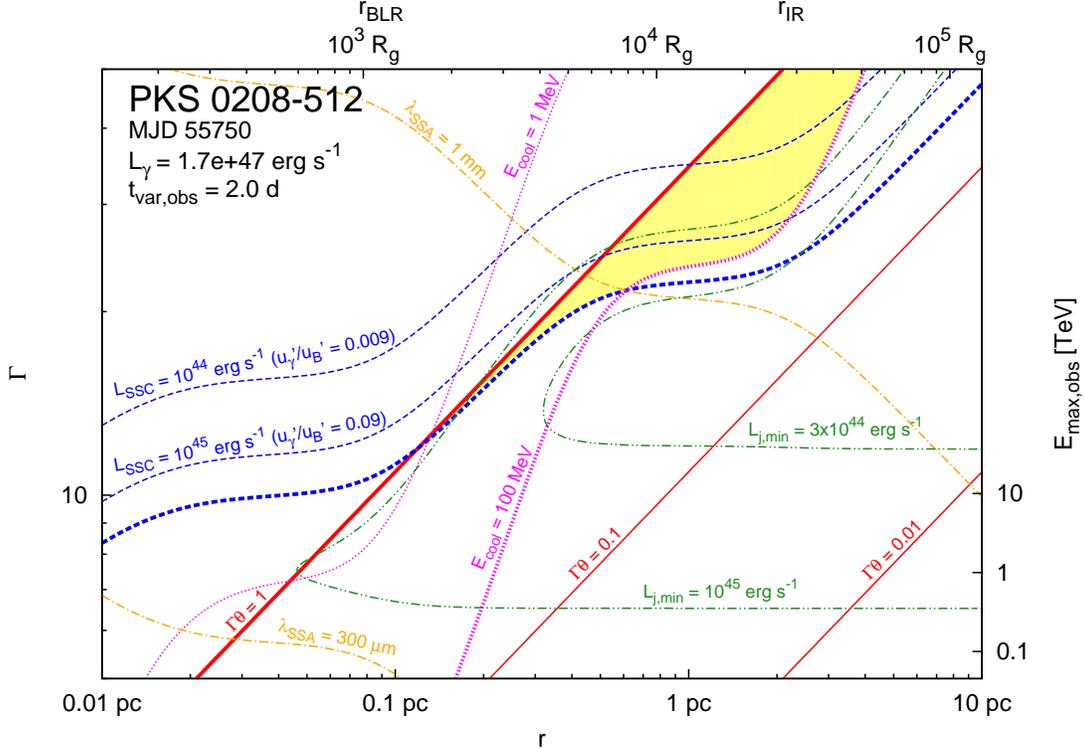}
\caption{Parameter space of $r$ and $\Gamma$ for the major flare of PKS~0208-512 that peaked at MJD~55750. See Fig. \ref{fig_3c454.3_55520} for detailed description.}
\label{fig_pks0208}
\end{figure*}

\cite{2011A&A...534A..86T} modeled the broad-band SED of PKS~1222+216 for this particular event, considering three scenarios: a single compact emitting region for both VHE and GeV emission, separate emitting regions located outside the BLR, and separate emitting regions with the GeV radiation produced within the BLR.
For the GeV emitting regions, they adopted a Lorentz factor $\Gamma = 10$ and a Doppler factor $\mathcal{D} \simeq 20$.
However, because they fixed the jet opening angle, at different distances they adopted different radii for the emitting regions, corresponding to different variability time scales.
For the GeV emitting region located within the BLR, their model predicts a variability time scale of $\simeq 10\;{\rm h}$, and for the region located outside the BLR, it predicts a variability time scale of $\simeq 3\;{\rm d}$.
When using these time scales, and a Doppler-to-Lorentz factor ratio of $\mathcal{D}/\Gamma = 2$, our constraints are entirely consistent with the model parameters adopted by \cite{2011A&A...534A..86T} in either scenario.

A characteristic feature of all the models of \cite{2011A&A...534A..86T} is that the magnetic component of the jet power is strongly dominated by the particle component, which in turn is dominated by protons.
However, considering only the electrons, they predict that $u_{\rm e}'/u_{\rm B}' \simeq 6$.
Even if only a moderate fraction of the energy of electrons can power the gamma-ray emission, their model is consistent with our result that $u_\gamma'/u_{\rm B}' \lesssim 10$.
We find that at moderate values of the Lorentz factor $\Gamma$ the jet can only be weakly magnetized.
If the extremely rapid VHE variability is due to processes powered by relativistic magnetic reconnection \citep{2012MNRAS.425.2519N,2013MNRAS.431..355G}, this requires a high jet magnetization, which is possible at the $\sim {\rm pc}$ scale, but only for very high Lorentz factors ($\Gamma \gtrsim 40$).
Alternatively, the required regions of very high magnetization may only occupy a small fraction of the jet cross-section.

\subsection{PKS~0208-512 at MJD~55750}
\label{sec_pks0208}

PKS~0208-512 ($z = 1.003$, $d_{\rm L} \simeq 6.7\;{\rm Gpc}$) showed several gamma-ray flares of moderate luminosity, which were studied in detail by \cite{2013ApJ...763L..11C}.
What is interesting about these flares is that they show significantly variable Compton dominance parameter.
Here we discuss the constraints on the parameters of one of the brightest gamma-ray flares produced by this source, peaking around MJD~55750.
Preliminary results for this event were presented in \cite{2013ApJ...771L..25C}.
Following that work, we adopt the following parameter values: $L_\gamma \simeq 1.7\times 10^{47}\;{\rm erg\,s^{-1}}$, $t_{\rm var,obs} \simeq 2\;{\rm d}$, $q = L_\gamma/L_{\rm syn} \simeq 3.3$, $L_X \simeq 3.5\times 10^{45}\;{\rm erg\,s^{-1}}$, $\alpha \simeq 0.7$, $L_{\rm d} \simeq 8\times 10^{45}\;{\rm erg\,s^{-1}}$, $\xi \simeq 0.1$, $r_{\rm BLR} \simeq 0.09\;{\rm pc}$, and $r_{\rm IR} \simeq 2.2\;{\rm pc}$.
While \cite{2013ApJ...771L..25C} adopted $\mathcal{D}/\Gamma \simeq 1.4$, here we will use $\mathcal{D}/\Gamma = 1$ as we do for all other sources.
We also adopt a black hole mass of $M_{\rm BH} \simeq 1.6\times 10^9\;M_\sun$ \citep{2004ApJ...602..103F}.

Our constraints for the gamma-ray flare in PKS~0208-512 are shown in Figure \ref{fig_pks0208}.
The marginal solution is uncertain in this case, because the $L_{\rm SSC}$ constraint is almost tangent to the collimation constraint, nevertheless, we adopt $r_{\rm min} \simeq 0.2\;{\rm pc}$ and $\Gamma_{\rm min} \simeq 15$.
With a relatively massive black hole, we have $r_{\rm min} \simeq 2500 R_{\rm g}$.
The predictions of this solution are: $\lambda_{\rm SSA,obs} \simeq 0.65\;{\rm mm}$, $u_\gamma'/u_{\rm B}' \simeq 0.26$, and $L_{\rm j,min} \simeq 0.92\times 10^{45}\;{\rm erg\,s^{-1}} \simeq 0.12L_{\rm d}$.
The cooling constraint, which was not considered by \cite{2013ApJ...771L..25C}, is rather strong, indicating that the jet cannot be strongly collimated, with $\Gamma\theta \gtrsim 0.3$.
This means that the emitting region must be located beyond $r_{\rm BLR}$, and possibly close to $r_{\rm IR}$.

\begin{figure*}[ht]
\includegraphics[width=0.5\textwidth]{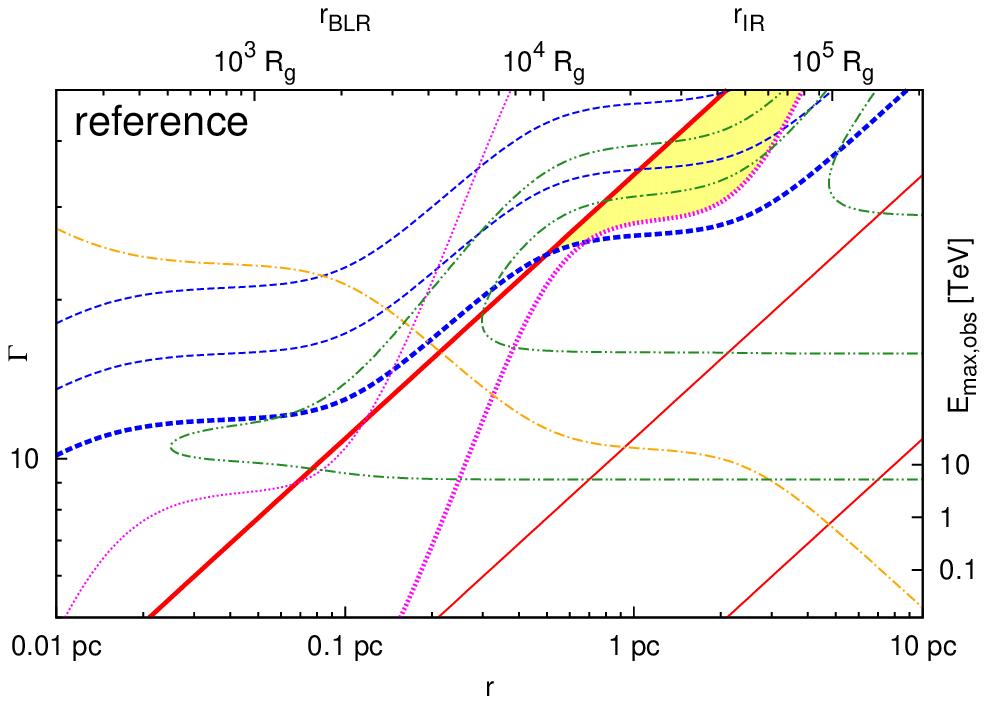}
\includegraphics[width=0.5\textwidth]{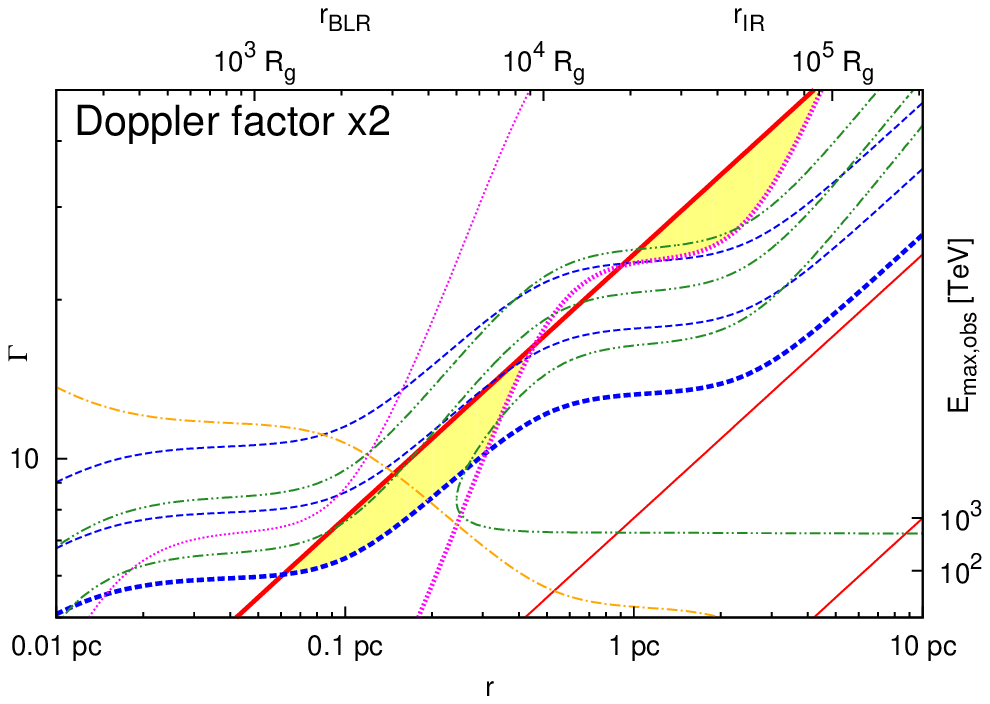}
\includegraphics[width=0.5\textwidth]{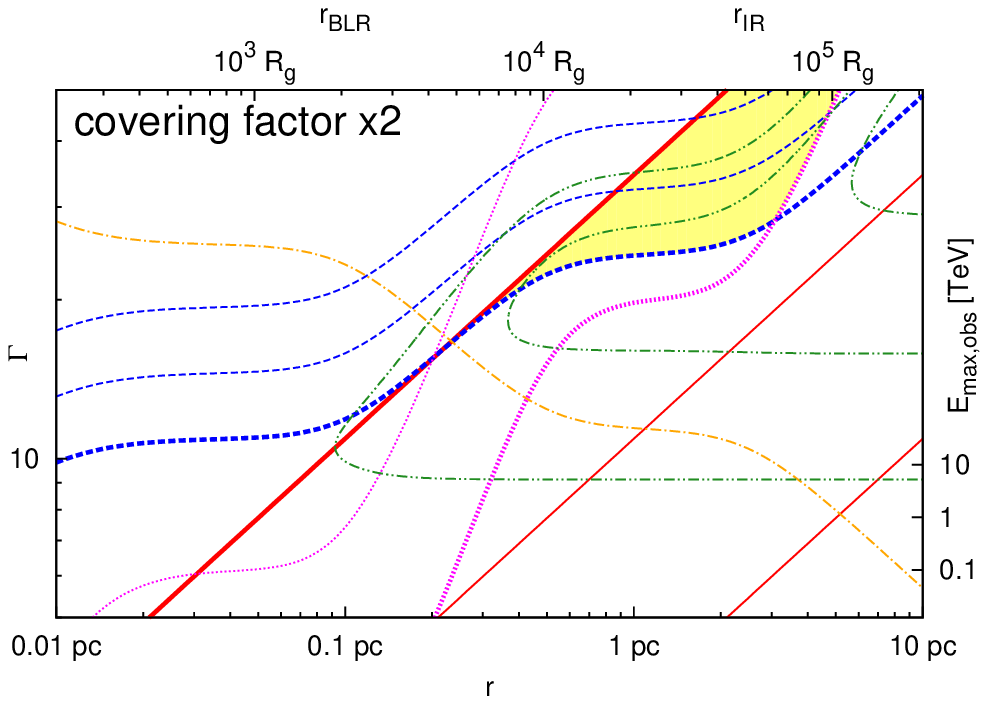}
\includegraphics[width=0.5\textwidth]{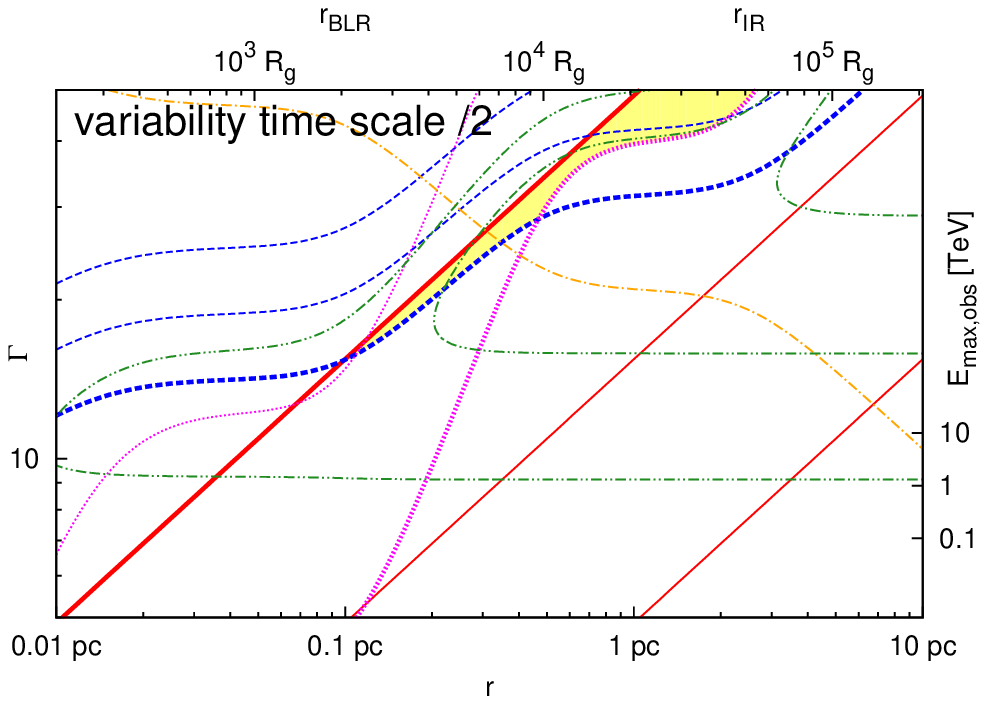}
\caption{Illustration of the sensitivity of our constraints to the assumptions on the Doppler-to-Lorentz factor ratio $\mathcal{D}/\Gamma$, the external radiation source covering factor $\xi$, and the observed variability time scale $t_{\rm var,obs}$. See Fig. \ref{fig_3c454.3_55520} for detailed description.}
\label{fig_sens}
\end{figure*}

\section{Sensitivity to assumptions}
\label{sec_sens}

There are parameters in our constraints, as in every model of blazar emission, that may not be well determined from observations.
In practice, even an informed choice of the values of these parameters is to some degree an arbitrary assumption.
In this section, we will discuss the sensitivity of our constraints to three such parameters: the Doppler-to-Lorentz factor ratio $\mathcal{D}/\Gamma$, the covering factor of external radiation sources $\xi_{\rm ext}$ (where `ext' stands for either BLR or IR), and the observed variability time scale $t_{\rm var,obs}$.

In fact, each of these three parameters can be estimated observationally to some degree.
As we mentioned at the beginning of Section \ref{sec_constr}, $\mathcal{D}$ and $\Gamma_{\rm j}$ can be deduced independently from the pc-scale jet kinematics probed by VLBI radio observations \citep{2005AJ....130.1418J,2009A&A...494..527H}.
However, in some cases it is found that $\mathcal{D}/\Gamma_{\rm j} \ll 1$, which is inconsistent with a blazar (PKS~1222+216, see Section \ref{sec_pks1222} and references therein).
The covering factors $\xi_{\rm ext}$ can be estimated in those sources where both the accretion disk continuum and broad emission lines or the infrared thermal component can be observed directly (PKS~1222+216, see Section \ref{sec_pks1222}), however, the geometry of external radiation sources (spherical --- planar) is uncertain, and it has a strong effect on the local energy densities \citep{2012arXiv1209.2291T,2013ApJ...779...68S}.
Thus, for most sources we adopted a fiducial value of $\xi_{\rm ext} \simeq 0.1$.
The variability time scale $t_{\rm var,obs}$ is a direct observable, however, it is a common situation that different values are adopted in independent studies of the same events (see Sections \ref{sec_3c454.3_55168}, \ref{sec_3c279}, \ref{sec_pks1510}).

In Figure \ref{fig_sens}, we show our constraints for 4 closely related fiducial models.
The reference model is calculated for $L_\gamma = 10^{48}\;{\rm erg\,s^{-1}}$, $t_{\rm var,obs} = 1\;{\rm d}$, $q = L_\gamma/L_{\rm syn} = 10$, $L_{\rm syn}/L_{\rm X} = 10$, $L_{\rm d} = 10^{46}\;{\rm erg\,s^{-1}}$, $\mathcal{D}/\Gamma = 1$, $\xi_{\rm BLR} = \xi_{\rm IR} = 0.1$, and $M_{\rm BH} = 10^9\,M_\sun$.
The second model differs from the reference model by having $\mathcal{D}/\Gamma \simeq 2$.
The third model differs from the reference model by having $\xi_{\rm BLR} = \xi_{\rm IR} = 0.2$.
Finally, the fourth model differs from the reference model by having $t_{\rm var,obs} = 12\,{\rm h}$.

The effect of increasing the Doppler-to-Lorentz factor ratio $\mathcal{D}/\Gamma$ is to significantly relax the SSC constraint, allowing for much lower values of $\Gamma$.
The collimation constraint is somewhat stronger, but the net effect of these two constraints is to decrease $r_{\rm min}$.
This can be understood from the fact that $\Gamma(r,\Gamma\theta) \propto (\mathcal{D}/\Gamma)^{-1/2}$ and $\Gamma(r,L_{\rm SSC}) \propto (\mathcal{D}/\Gamma)^{-1}$ (see Eqs. \ref{eq_constr_Gammatheta} and \ref{eq_constr_LSSC}).
The cooling constraint is affected only slightly, since $\Gamma(r,E_{\rm cool,obs}) \propto (\mathcal{D}/\Gamma)^{-1/4}$ (see Eq. \ref{eq_constr_Ecool}).
The relation between the `equipartition' parameter $u_\gamma'/u_{\rm B}'$ and the SSC constraint is independent of $\mathcal{D}/\Gamma$ (see Eq. \ref{eq_constr_ugammauB}), therefore lines of constant $L_{\rm SSC}$ correspond to the same values of $u_\gamma'/u_{\rm B}'$ as in the reference model.
The dependence of the intrinsic opacity constraint $\Gamma(E_{\rm max,obs})$ on $\mathcal{D}/\Gamma$ (see Eq. \ref{eq_constr_Emax}) is the same as that of the SSC constraint. Although the gradients of $E_{\rm max,obs}$ in the $(r,\Gamma)$ space are large, the value of $E_{\rm max,obs}$ for the marginal solution decreases only slightly.
The minimum jet power is also significantly relaxed, especially in the region dominated by the radiation energy density.
However, in the more relevant region dominated by the magnetic energy density, $\Gamma(r,L_{\rm j,B,min}) \propto (\mathcal{D}/\Gamma)^{-2/3}$ (see Eq. \ref{eq_constr_LjBmin}), and the lines of constant $L_{\rm j,min}$ are aligned roughly parallel to the lines of constant $L_{\rm SSC}$.
Because of steep gradients of $L_{\rm j,B,min}$ in the $(r, \Gamma)$ space, its value is very sensitive to the exact location within the allowed region.
Finally, the dependence of the SSA constraint $\Gamma(r,\nu_{\rm SSA})$ on $\mathcal{D}/\Gamma$ (see Eq. \ref{eq_constr_nuabsobs}) is the same as that of the SSC constraint, and the gradients of $\nu_{\rm SSA,obs}$ in the $(r,\Gamma)$ space are very small.
Therefore, the predicted SSA characteristic frequency for the marginal solution will be only weakly affected.
We conclude that while the allowed parameter space region for higher $\mathcal{D}/\Gamma$ is significantly extended towards lower values of $r$ and $\Gamma$, most parameter values corresponding to the marginal solution ($r_{\rm min}$, $\Gamma_{\rm min}$) are not very sensitive to the choice of $\mathcal{D}/\Gamma$.

The effect of increasing the covering factor $\xi \equiv \xi_{\rm BLR} = \xi_{\rm IR}$ is relatively minor.
Our constraints scale with $\xi$ like: $\Gamma(r,\Gamma\theta) \propto \xi^0$, $\Gamma(r,L_{\rm SSC}) \propto \xi^{-1/8}$, $\Gamma(r,E_{\rm cool}) \propto \xi^{-1/2}$, $\Gamma(r,\nu_{\rm SSA,obs}) \propto \xi^{1/8}$, $\Gamma(E_{\rm max,obs}) \propto \xi^0$, $\Gamma(L_{\rm j,\gamma,min}) \propto \xi^0$, and $\Gamma(r,L_{\rm j,B,min}) \propto \xi^{-1/6}$.
The cooling constraint is moderately relaxed, extending the allowed parameter space region towards higher values of $r$.
Other scalings are very weak, and therefore we conclude that the choice of $\xi$ is not critical in our analysis.

The effect of decreasing the observed variability time scale $t_{\rm var,obs}$ is quite significant.
Our constraints scale with $t_{\rm var,obs}$ like: $\Gamma(r,\Gamma\theta) \propto t_{\rm var,obs}^{-1/2}$, $\Gamma(r,L_{\rm SSC}) \propto t_{\rm var,obs}^{-1/4}$, $\Gamma(r,E_{\rm cool}) \propto t_{\rm var,obs}^{-1/2}$, $\Gamma(r,\nu_{\rm SSA,obs}) \propto t_{\rm var,obs}^{-1}$, $\Gamma(E_{\rm max,obs}) \propto t_{\rm var,obs}^{-1/(4+2\alpha)}$, $\Gamma(L_{\rm j,\gamma,min}) \propto t_{\rm var}^0$, and $\Gamma(r,L_{\rm j,B,min}) \propto t_{\rm var,obs}^{-1/3}$.
The allowed parameter space region is shifted towards smaller values of $r$ due to relaxed collimation constraint and tighter cooling constraint, and the SSA is noticeably stronger, but other parameters are not strongly affected.

In summary, the uncertainty in the Doppler-to-Lorentz factor ratio is the most significant unknown in our model, but the general conclusions that we draw for each case analyzed in Section \ref{sec_cases} are securely robust.

\begin{table*}[ht]
\small
\centering
\caption{Parameters of our constraints, the marginal solutions (minimum distances), and the maximum distances for all blazar flares studied in Section \ref{sec_cases}.}
\vskip 1ex
\label{tab_param}
\begin{tabular}{cccccccc}
\hline\hline
object
& 3C
& 3C
& AO
& 3C
& PKS
& PKS
& PKS
\\
& 454.3
& 454.3
& 0235+164
& 279
& 1510-089
& 1222+216
& 0208-512
\\
MJD
& 55520
& 55168
& 54760
& 54880
& 54948
& 55366
& 55750
\\
& (a,b,c)
& (d,e,f)
& (g,h)
& (i,j,k)
& (l,m,n)
& (o,p)
& (q,r)
\\
\hline
$L_\gamma\;[10^{48}\,\rm erg\,s^{-1}]$
& 47
& 8.4
& 0.67
& 0.26
& 0.54
& 1
& 0.17
\\
$t_{\rm var}\;[\rm d]$
& 0.36
& 1
& 3
& 1.5
& 0.9
& 1
& 2
\\
$q = L_\gamma/L_{\rm syn}$
& 30
& 14
& 4
& 7.5
& 100
& 100
& 3.3
\\
$L_\gamma/L_X$
& 300
& 140
& 24
& 69
& 1000
& 1000
& 49
\\
$L_{\rm d}\;[10^{46}\,\rm erg\,s^{-1}]$
& 6.75
& 6.75
& 0.4
& 0.2
& 0.5
& 5
& 0.8
\\
$M_{\rm bh}\;[10^8\,M_\sun]$
& 5
& 5
& 4
& 5
& 4
& 6
& 16
\\
\hline
$r_{\rm min}\;[\rm pc]$
& 0.16
& 0.17
& 0.65
& 0.62
& 0.37
& 0.18
& 0.2
\\
$r_{\rm min}/R_{\rm g}\;[10^3]$
& 6.6
& 7
& 33
& 26
& 19
& 6.2
& 2.5
\\
$r_{\rm min}/r_{\rm BLR}$
& 0.62
& 0.65
& 10
& 14
& 5.3
& 0.8
& 2.2
\\
$\Gamma_{\rm min}$
& 30
& 19
& 22
& 27
& 26
& 17
& 15
\\
$\lambda_{\rm SSA,obs}\;[\rm mm]$
& 0.125
& 0.215
& 0.92
& 1.03
& 1.4
& 0.76
& 0.65
\\
$u_\gamma'/u_{\rm B}'$
& 3.3
& 1.6
& 0.7
& 0.3
& 12
& 11
& 0.26
\\
$L_{\rm j,min}\;[10^{45}\,\rm erg\,s^{-1}]$
& 17
& 10
& 0.85
& 0.4
& 0.22
& 0.95
& 0.92
\\
$L_{\rm j,M11}\;[10^{45}\,\rm erg\,s^{-1}]$ (s)
& 2
& 2
& 0.2
& 0.9
& 0.3
& 0.8
& ---
\\
\hline
$r_{\rm max}\;{\rm [pc]}$ (*)
& 0.8
& 8.5
& 3.4
& 1.7
& 2.4
& 10.7
& 4
\\
$r_{\rm max}/r_{\rm min}$
& 5
& 50
& 5.2
& 2.8
& 6.4
& 59
& 20
\\
$r_{\rm max}/r_{\rm IR}$
& 0.12
& 1.3
& 2.1
& 1.5
& 1.4
& 1.9
& 1.8
\\
\hline\hline
\end{tabular}
\vskip 1ex
References:
a~---~\cite{2011ApJ...733L..26A};
b~---~\cite{2011ApJ...736L..38V};
c~---~\cite{2012ApJ...758...72W};
d~---~\cite{2010ApJ...721.1383A};
e~---~\cite{2011MNRAS.410..368B};
f~---~\cite{2010ApJ...716L.170P};
g~---~\cite{2012ApJ...751..159A};
h~---~\cite{2011ApJ...735L..10A};
i~---~\cite{2010Natur.463..919A};
j~---~\cite{2012ApJ...754..114H};
k~---~\cite{2014ApJ...782...82D};
l~---~\cite{2010ApJ...721.1425A};
m~---~\cite{2011A&A...529A.145D};
n~---~\cite{2010ApJ...710L.126M};
o~---~\cite{2011ApJ...733...19T};
p~---~\cite{2011A&A...534A..86T};
q~---~\cite{2013ApJ...763L..11C};
r~---~\cite{2013ApJ...771L..25C};
s~---~\cite{2011ApJ...740...98M}.
\\
(*) Calculated for $\Gamma_{\rm max} = 50$, and in the case of 3C~279 for $\Gamma_{\rm max} \simeq 46$.
\end{table*}

\section{Discussion}
\label{sec_disc}

We have demonstrated that it is possible to significantly constrain the parameter space of distance from the central SMBH $r$ and Lorentz factor $\Gamma$ of emitting regions responsible for bright gamma-ray flares of luminous blazars in the framework of the ERC mechanism, using 5 direct observables: gamma-ray luminosity $L_\gamma$, gamma-ray variability time scale $t_{\rm var,obs}$, synchrotron luminosity $L_{\rm syn}$, X-ray luminosity $L_{\rm X}$, and accretion disk luminosity $L_{\rm d}$.
A combination of the collimation constraint ($\Gamma\theta \lesssim 1$), the SSC constraint ($L_{\rm SSC} \lesssim L_{\rm X}$), and the cooling constraint ($E_{\rm cool,obs} \lesssim 100\;{\rm MeV}$) defines a parameter space region such that for each value of $\Gamma > \Gamma_{\rm min}$, the range of $r$ is limited to factor $\sim 2-10$.
This is a significant improvement over previous studies, which are typically limited to deciding between the BLR and IR regions, with $r_{\rm IR}/r_{\rm BLR} \sim 30$ \citep[\eg,][]{2009ApJ...704...38S,2012ApJ...758L..15D,2013MNRAS.431..824B}.
Moreover, we evaluate the effect on our results of the most uncertain parameters like Doppler-to-Lorentz factor ratio $\mathcal{D}/\Gamma$, or covering factor $\xi$ of external radiation sources.
Further progress is possible with improved multiwavelength observations of blazars, if they can be used to securely pinpoint the synchrotron self-absorption frequency $\nu_{\rm SSA,obs}$.

\subsection{Collimation parameter}
\label{sec_disc_coll_par}

While we have imposed an upper limit on the collimation parameter $\Gamma\theta \lesssim 1$, the SSC and cooling constraints provide a firm lower limit.
In some analyzed cases (Figure \ref{fig_3c279}), this limit is as strong as $\Gamma\theta \gtrsim 0.7$.
In other cases (Figure \ref{fig_pks1222}), values of $\Gamma\theta \simeq 0.1$ can be obtained only for Lorentz factors $\Gamma \gtrsim 25$.
Such tight lower limits may be in conflict with VLBI radio observations that imply significantly tighter upper limits, with $\Gamma_{\rm j}\theta_{\rm j} \lesssim 0.3$ \citep{2009A&A...507L..33P,2013A&A...558A.144C}.
However, these radio observations probe the jet geometry at many-pc scales, and it is not clear whether these results are relevant for pc-scale jets.
Also, the Lorentz factor $\Gamma$ of the emitting region may be larger than the jet Lorentz factor $\Gamma_{\rm j}$.
In any case, we can securely conclude that \emph{very narrow opening angles of the gamma-ray emitting regions are excluded by the SSC and cooling constraints}.
This makes any model of energy dissipation in jets which operates on a small fraction of the jet cross-section, in particular \emph{reconfinement shocks leading to very narrow nozzles} \citep[\eg,][]{2009ApJ...699.1274B}, inconsistent with the ERC scenario.
This also challenges models of strongly structured jets, e.g. the spine-sheath models \citep{2005A&A...432..401G}, or models involving strongly localized dissipation sites, e.g. minijets \citep{2009MNRAS.395L..29G}, unless they can be distributed uniformly across a large fraction of the jet cross-section.
While these models can still explain the most extreme modes of blazar variability, in particular the sub-hour very high energy gamma-ray flares \citep{2011ApJ...730L...8A}, they may not be responsible for the bulk of the gamma-ray emission of blazars.

\subsection{Marginal solutions}
\label{sec_disc_marg_sol}

The intersection between the collimation constraint and the SSC constraint defines the marginal solution ($r_{\rm min}$, $\Gamma_{\rm min}$), which sets firm lower limits on both $r$ and $\Gamma$. One can derive the marginal solution from Equations (\ref{eq_constr_Gammatheta}) and (\ref{eq_constr_LSSC}):
\bea
\label{eq_rmin}
r_{\rm min} &\simeq&
\frac{ct_{\rm var,obs}}{(1+z)}
\left[\frac{3}{4}\left(\frac{g_{\rm SSC}}{g_{\rm ERC}}\right)\left(\frac{L_{\rm syn}}{L_{\rm X}}\right)\left(\frac{L_\gamma}{\zeta(r_{\rm min})L_{\rm d}}\right)\right]^{1/2}
\times\nonumber\\&&
\left(\frac{\mathcal{D}}{\Gamma}\right)^{-2}
\,, \\
\Gamma_{\rm min} &\simeq&
\left[\frac{3}{4}\left(\frac{g_{\rm SSC}}{g_{\rm ERC}}\right)\left(\frac{L_{\rm syn}}{L_{\rm X}}\right)\left(\frac{L_\gamma}{\zeta(r_{\rm min})L_{\rm d}}\right)\right]^{1/4}
\times\nonumber\\&&
\left(\frac{\mathcal{D}}{\Gamma}\right)^{-3/2}
\,.
\eea
Because of the dependence of $\zeta$ on $r$, Eq. (\ref{eq_rmin}) is not explicit, but the solutions discussed below are calculated self-consistently.
One can see that the minimum distance scale $r_{\rm min}$ is proportional to the observed variability time scale $t_{\rm var,obs}$. Both $r_{\rm min}$ and $\Gamma_{\rm min}$ depend strongly on the Doppler-to-Lorentz ratio, and they are weak functions of the broad-band SED shape.
The marginal solutions for the cases analyzed in Section \ref{sec_cases} are listed in Table \ref{tab_param}.
Even with this very small sample, we can point to some general trends and differences.
The minimum distance ranges between $0.16 \lesssim r_{\rm min}\,[{\rm pc}] \lesssim 0.65$.
In terms of gravitational radii, the range is $2500 \lesssim r_{\rm min}/R_{\rm g} \lesssim 33000$, which is much wider than the spread of black hole mass estimates for the 6 analyzed blazars --- $4\times 10^8 \lesssim M_{\rm BH}/M_\sun \lesssim 1.6\times 10^9$.
In terms of the BLR radii, the range is $0.62 \lesssim r_{\rm min}/r_{\rm BLR} \lesssim 14$.
Interestingly, the range of absolute values of $r_{\rm min}$ is much narrower than the ranges of relative values of $r_{\rm min}/R_{\rm g}$ and $r_{\rm min}/r_{\rm BLR}$.
Flares with relatively large $r_{\rm min}$ happen to be both long and faint.
The minimum Lorentz factor ranges between $15 \lesssim \Gamma_{\rm min} \lesssim 30$.
It does not show an obvious trend with the gamma-ray luminosity $L_\gamma$ or with the time scale $t_{\rm var,obs}$.

The energy density ratio of the gamma-ray radiation to the magnetic fields for the marginal solution is given by (cf. Equation \ref{eq_constr_ugammauB}):
\be
\frac{u_\gamma'}{u_{\rm B}'} \simeq \frac{L_\gamma L_{\rm X}}{g_{\rm SSC}L_{\rm syn}^2}\,.
\ee
One can see that it depends only on the broad-band SED shape.
From Table \ref{tab_param}, we find that it ranges between $0.26 \lesssim u_\gamma'/u_{\rm B}' \lesssim 12$.
Values lower by about order of magnitude are possible for other solutions, which also have lower values of $L_{\rm SSC}$.
The energy density ratio generally increases with the Compton dominance parameter $q$.
The gamma-ray radiation density $u_\gamma'$ closely probes the high-energy end of the electron population, and provides a lower limit on the total electron energy density $u_{\rm e}'$.
Assuming very roughly that $3 \lesssim u_{\rm e}'/u_\gamma' \lesssim 10$, we can expect that $u_{\rm e}'/u_{\rm B}' \sim 0.08 - 120$.
In this sense, \emph{our constraints are not in conflict with the equipartition condition $u_{\rm e}'/u_{\rm B}' \simeq 1$}, which is sometimes imposed on blazar models \citep[\eg,][]{2009ApJ...703.1168B,2014ApJ...782...82D}.
This also indicates that \emph{(sub-)pc scale jets are at most moderately magnetized.}
Very high magnetization values would require violating the jet collimation constraint, i.e. $\Gamma\theta > 1$.

The minimum required jet power for the marginal solution is given by (cf. Equations \ref{eq_constr_Ljmin1} and \ref{eq_constr_Ljmin2}):
\bea
L_{\rm j,min} &=&
\frac{L_\gamma}{4}
\left[\frac{3}{4}\left(\frac{g_{\rm SSC}}{g_{\rm ERC}}\right)\left(\frac{L_{\rm syn}}{L_{\rm X}}\right)\left(\frac{L_\gamma}{\zeta(r_{\rm min})L_{\rm d}}\right)\right]^{-1/2}\times
\nonumber\\&&
\left(\frac{\mathcal{D}}{\Gamma}\right)^{-1}
\left(1+\frac{u_{\rm B}'}{u_\gamma'}\right)
\,.
\eea
One can see that it depends primarily on the gamma-ray luminosity, relatively weakly on the Doppler-to-Lorentz ratio, and to some degree also on the broad-band SED shape.
Our estimates of the minimum jet power for the analyzed cases (Table \ref{tab_param}) range between $2.2\times 10^{44} \lesssim L_{\rm j,min}\,[{\rm erg\,s^{-1}}] \lesssim 1.7\times 10^{46}$, which is significantly narrower than the range of apparent gamma-ray luminosities $1.7\times 10^{47} \lesssim L_\gamma\,[{\rm erg\,s^{-1}}] \lesssim  4.7\times 10^{49}$.
In terms of the accretion disk luminosity, we find $0.019 \lesssim L_{\rm j,min}/L_{\rm d} \lesssim 0.25$.
There is a trend for this ratio to be higher for lower Compton dominance $q$ (and higher jet magnetization).
For 5 blazars (excluding PKS~0208-512), we compare $L_{\rm j,min}$ with the estimates $L_{\rm j,M11}$ of total jet power by \cite{2011ApJ...740...98M}.
We find that in many cases our lower limits significantly exceed $L_{\rm j,M11}$, with $0.44 \lesssim L_{\rm j,min}/L_{\rm j,M11} \lesssim 8.5$.
Since our estimates do not take into account the contributions from cold/warm electrons and protons, \emph{the total jet powers required to power the observed gamma-ray flares may be comparable to, or even exceed, the accretion disk luminosity} \citep[in agreement with][]{2009MNRAS.399.2041G}, and they are certain to be significantly higher than the estimates of \cite{2011ApJ...740...98M}.
This indicates that the total jet powers in blazars are strongly variable, and that the values estimated from energetics of the brightest gamma-ray flares (this work) can exceed by more than order of magnitude higher the average values inferred from the low-frequency (300 MHz) radio luminosity \citep{2011ApJ...740...98M}.

The synchrotron self-absorption (SSA) wavelength for the marginal solutions ranges between $0.125 \lesssim \lambda_{\rm SSA,obs}\;{\rm [mm]} \lesssim 1.4$.
For other allowed solutions $\lambda_{\rm SSA}$ will be somewhat larger.
The SSA threshold appears to be better correlated with the gamma-ray luminosity $L_\gamma$ than with the observed variability time scale $t_{\rm var,obs}$.
For 5 events with marginal $\lambda_{\rm SSA,obs} > 0.5\;{\rm mm}$, a fairly close correlation between the gamma rays and the mm data can be expected \citep{2008ApJ...675...71S}.
However, for the bright flares of 3C~454.3, where marginal $\lambda_{\rm SSA,obs} \lesssim 0.2\;{\rm mm}$, we expect that the mm signal should be significantly delayed with respect to the gamma-ray signal.
The gamma-ray emitting regions for the analyzed events cannot be transparent at the $7\;{\rm mm}$ wavelength.
Because of the weak dependence of $\lambda_{\rm SSA,obs}$ on either $r$ or $\Gamma$, \emph{SSA can potentially provide very strong additional constraints on the parameters of gamma-ray emitting regions in blazars.}

The intrinsic gamma-ray opacity does not provide a significant constraint in the analyzed cases, with $\Gamma(E_{\rm max,obs} = 100\;{\rm GeV}) \lesssim 10$.
In every analyzed case, the SSC constraint gives a stronger lower limit on $\Gamma$, as first noted by \cite{2010ApJ...721.1383A}.

\subsection{Maximum distance scale}
\label{sec_disc_rmax}

For a given value of the Lorentz factor $\Gamma$, the maximum distance $r_{\rm max}(\Gamma)$ is determined either by the SSC constraint, or by the cooling constraint.
Eventually, at some $\Gamma_{\rm max}$ there is a solution where the cooling constraint crosses the collimation constraint, which gives an absolute upper limit $r_{\rm max}(\Gamma_{\rm max})$.
However, the values of $\Gamma_{\rm max}$ can be extremely high ($\Gamma_{\rm max} \gg 50$), especially for sources with high accretion disk luminosity $L_{\rm d}$ (3C~454.3 and PKS~1222+216), for which the cooling constraint is relatively weak.
Therefore, the effective maximum distance scale depends on how high values of $\Gamma$ one would accept.\footnote{VLBI observations indicate that jet Lorentz factors for luminous blazars are $\Gamma_{\rm j} \lesssim 35$ \citep{2009A&A...494..527H}. Moreover, $\Gamma_{\rm j} \gg 15$ would contradict the blazar beaming statistics \citep[\eg,][]{2010ApJ...721.1383A}. However, in this work we explicitly allow for $\Gamma > \Gamma_{\rm j}$.}
For a rather high $\Gamma_{\rm max} = 50$ ($\Gamma_{\rm max} \simeq 46$ in the case of 3C~279), we obtain $0.8 \lesssim r_{\rm max}\;{\rm [pc]} \lesssim 10.7$ (see Table \ref{tab_param}).
In terms of the IR radii, the range is $0.12 \lesssim r_{\rm max}/r_{\rm IR} \lesssim 2.1$.
The ratio of maximum to minimum distances is in the range $2.8 \lesssim r_{\rm max}/r_{\rm min} \lesssim 59$.
The distance scale is best constrained for the flare in 3C~279, which is characterized by the lowest value of $L_{\rm d}$.

If the cooling constraint can be relaxed due to the swinging motion of the emitting region, we can still place significant limits on the far-dissipation scenario by using solely the SSC constraint.
In most analyzed cases, locating the gamma-ray emitting regions at $r \simeq 10\;{\rm pc}$ would require $\Gamma > 50$.

\subsection{Limits to the ERC model}
\label{sec_disc_limerc}

In Section \ref{sec_ao0235}, we discussed the tension between the constraints imposed by the ERC model and the far dissipation ($\sim 10\;{\rm pc}$) scenarios motivated by the observed gamma-ray/mm-radio connection.
We showed that the SSC constraint requires very high Lorentz factors, $\Gamma \gtrsim 50$, in order for gamma-ray flares with variability time scale of $\sim 1\;{\rm d}$ to be produced at the distance scale of $\sim 10\;{\rm pc}$.
These solutions are also characterized by inefficient electron cooling ($E_{\rm cool,obs} \gg 100\;{\rm MeV}$), which would result in strongly asymmetric gamma-ray light curves with long flux-decay time scales, unless there are fast variations in the local Doppler factor.
Alternative sources of external radiation at large distance scales were proposed as a way around these problems.
In Appendix \ref{sec_disc_far_diss}, we discuss two such ideas --- spine-sheath models \citep[\eg,][]{2010ApJ...710L.126M}, and extended broad-line regions \citep{2011A&A...532A.146L}.

While far less popular than the ERC model, the SSC model is still being considered when modeling FSRQ blazars \citep[\eg,][]{2010ApJ...714L.303F,2012MNRAS.420...84Z,2012MNRAS.424..789C}.
It was suggested that the SSC model is most relevant for FSRQs with relatively low kinetic jet power \citep{2012ApJ...752L...4M}.
Such models can be characterized by two conditions: $L_{\rm SSC} = L_\gamma$ and $L_{\rm ERC} < L_{\rm SSC}$ (one should note that in this case $L_{\rm ERC}$ may be suppressed, being strongly in the Klein-Nishina regime due to higher electron energies). With minor modifications, we can use our constraints to identify the parameter space region where these conditions can be satisfied. Our SSC constraint (Eq. \ref{eq_constr_LSSC}) is more generally a constraint on the luminosity ratio $L_{\rm SSC}/L_{\rm ERC}$, which increases systematically with decreasing $\Gamma$. One can extrapolate from the lines of constant $L_{\rm SSC}$ shown on Figures \ref{fig_3c454.3_55520} - \ref{fig_pks0208} ($L_{\rm SSC}/L_{\rm ERC} = L_{\rm SSC}/L_\gamma \ll 1$) to the case of $L_{\rm SSC}/L_{\rm ERC} = L_\gamma/L_{\rm ERC} > 1$ corresponding to moderate and low Lorentz factors, $\Gamma \lesssim 10$. The parameter space of the SSC model is clearly separated from the parameter space of the ERC model. At distances of $\sim 10\;{\rm pc}$, SSC model may be favored over the ERC model, the latter requiring extreme values of $\Gamma$.

The jet collimation constraint is the same for the ERC and SSC models, as it does not depend on any kind of luminosity.
Because of the lower Lorentz factor characterizing the SSC model, it corresponds to very strong jet collimation, with $\Gamma\theta \lesssim 0.1$, especially at larger distances. \emph{An SSC model operating at the distance scale of $10\;{\rm pc}$ requires significant jet recollimation or sharp jet substructure.}
Other constraints are distance independent, as they no longer depend on the distribution of external radiation fields.
According to Eq. (\ref{eq_constr_ugammauB}), the `equipartition' parameter is $u_{\rm B}'/u_\gamma' \simeq g_{\rm SSC}/q^2 \ll 1$, therefore the SSC model implies a strongly particle-dominated emitting region \citep{2009ApJ...704...38S}.
The minimum required jet power, dominated by the radiative component, is comparable to or slightly larger than that in the ERC model.

There are additional very strong constraints on the SSC model from the observed broad-band SEDs of luminous blazars.
While being very successful in explaining the emission of low-luminosity HBL blazars, SSC models can have serious difficulties in matching the observed SEDs of FSRQs \citep[\eg,][]{2012cosp...39..848J}.
In order to match the characteristic frequencies of the two main spectral components, SSC models typically require very low magnetic field strength and high average electron random Lorentz factor, which independently suggests a particle-dominated emitting region.
A more detailed analysis of the spectral constraints on the SSC model is beyond the scope of this work.

\section{Conclusions}
\label{sec_conc}

We investigated several constraints on the location $r$ and the Lorentz factor $\Gamma$ of gamma-ray emitting regions in the jets of luminous blazars, assuming that the gamma-ray emission is produced by the external radiation Comptonization (ERC) mechanism.
In Section \ref{sec_constr}, we defined 4 such constraints, based on: collimation parameter $\Gamma\theta$, synchrotron self-Compton (SSC) luminosity $L_{\rm SSC}$, observed photon energy corresponding to efficient cooling threshold $E_{\rm cool,obs}$, and maximum photon energy $E_{\rm max,obs}$ due to intrinsic gamma-ray opacity.
In Section \ref{sec_pred}, we also considered specific predictions for given $(r,\Gamma)$ --- synchrotron self-absorption (SSA) frequency $\nu_{\rm SSA,obs}$, and minimum jet power $L_{\rm j,min}$ including only contributions from high-energy electrons and magnetic field.
In practical application, these constraints require 5 direct observables --- gamma-ray luminosity $L_\gamma$, gamma-ray variability time scale $t_{\rm var,obs}$, synchrotron luminosity $L_{\rm syn}$ (or Compton dominance parameter $q = L_\gamma/L_{\rm syn}$), X-ray luminosity $L_{\rm X}$, and accretion disk luminosity $L_{\rm d}$ --- and a small number of assumptions: Doppler-to-Lorentz factor ratio $\mathcal{D}/\Gamma$, and covering factors of external radiation sources $\xi_{\rm BLR}, \xi_{\rm IR}$.
The sensitivity of the constraints to the assumptions was evaluated in Section \ref{sec_sens}.
In Section \ref{sec_cases}, we applied these constraints to several well-known gamma-ray flares for which extensive multiwavelength data are available.
For each studied case, we plot the parameter space $(r,\Gamma)$ to illustrate our results (Figures \ref{fig_3c454.3_55520} -- \ref{fig_pks0208}).

We find that the most useful constraints on $r$ and $\Gamma$ can be derived from the combination of three conditions: $\Gamma\theta \lesssim 1$, $L_{\rm SSC} \lesssim L_{\rm X}$, and $E_{\rm cool,obs} \lesssim 100\;{\rm MeV}$.
They define a characteristic region in the parameter space anchored at the marginal solution $(r_{\rm min},\Gamma_{\rm min})$.
In the analyzed cases, we found that $0.16 \lesssim r_{\rm min}\;{\rm [pc]} \lesssim 0.65$ and $15 \lesssim \Gamma_{\rm min} \lesssim 30$.
Larger distances are possible only for higher Lorentz factors, but eventually they are limited by the cooling constraint.
The size of the allowed parameter space region is particularly small for sources with low accretion disk luminosity $L_{\rm d}$.

Our constraints challenge the far-dissipation scenarios inspired by the observed gamma-ray/mm connection.
As we show in Appendix \ref{sec_disc_gamma_mm}, light travel time effects can easily explain the temporal coincidence between gamma-ray flares and the radio/mm activity, even when the gamma-ray emitting region is located far upstream from the radio/mm core.
As we show in Appendix \ref{sec_disc_spine_sheath}, external radiation fields cannot be substituted at large distances by synchrotron radiation from a slower jet sheath.
However, as we discuss in Appendix \ref{sec_disc_ext_blr}, a scenario involving an extended broad-line region \citep{2011A&A...532A.146L} may provide an alternative source of external radiation.

The upper limit on $L_{\rm SSC}$ can be translated into a lower limit on the collimation parameter, $\Gamma\theta \gtrsim 0.1-0.7$, which means that dissipation cannot be limited to very compact jet substructures like reconfinement nozzles, spines, minijets, etc.
Our results support the idea that pc-scale blazar jets should be close to energy equipartition between the particle and magnetic components.

The intrinsic opacity constraint on the Lorentz factor is always weaker than the SSC constraint.
The synchrotron self-absorption constraint can significantly improve the determination of the parameters of gamma-ray emitting regions, if sufficient multiwavelength data can be collected, possibly resolving the degeneracy in the values of the Doppler and covering factors.

\acknowledgments 

We thank the anonymous referee for valuable comments on the manuscript, and Alan Marscher for discussions.
K.N. thanks the staff of the Nicolaus Copernicus Astronomical Center for their hospitality during the preparation of this manuscript.
This project was partly supported by NASA through the Fermi Guest Investigator program,
and by the Polish NCN through grant DEC-2011/01/B/ST9/04845.
K.N. was supported by NASA through Einstein Postdoctoral Fellowship grant number PF3-140112 awarded by the Chandra X-ray Center, which is operated by the Smithsonian Astrophysical Observatory for NASA under contract NAS8-03060.

\appendix

\section{External radiation distribution}
\label{sec_ext_rad}

\begin{figure}[ht]
\centering
\includegraphics[width=\columnwidth]{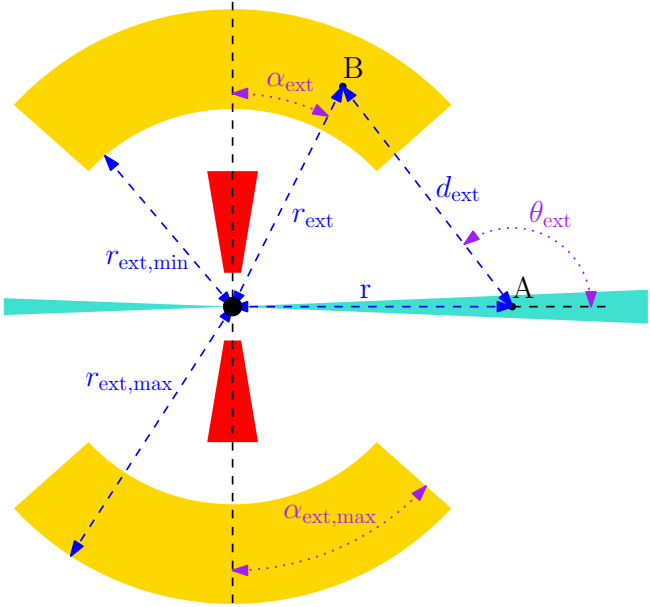}
\caption{Geometry of the external radiation emitting region adopted in this work for both the broad-line region and the dusty torus. See Appendix \ref{sec_ext_rad} for details.}
\label{fig_ext_rad}
\end{figure}

In this work we adopt a specific geometry of the broad-line region and the dusty torus where external radiation fields are produced (see Figure \ref{fig_ext_rad}).
Both regions are assumed to be symmetric with respect to the jet axis, and they span a distance range (measured from the SMBH) of $r_{\rm ext,min} \le r_{\rm ext} \le r_{\rm ext,max}$ and an equatorial angle range (measured from the accretion disk plane, perpendicular to the jet axis) of $-\alpha_{\rm ext,max} \le \alpha_{\rm ext} \le \alpha_{\rm ext,max}$.
The fraction of accretion disk radiation reprocessed over unit radius ${\rm d}r_{\rm ext}$ is assumed to scale like $\xi(r_{\rm ext}) \propto r_{\rm ext}^{-\beta_{\rm ext}}$, and it is normalized so that the effective covering factor is $\xi_{\rm ext} = \int_{r_{\rm ext,min}}^{r_{\rm ext,max}}\xi(r)\;{\rm d}r_{\rm ext}$.

This simple model has only 4 significant parameters: $r_{\rm ext,min}$, $\alpha_{\rm ext,max}$, $\xi_{\rm ext}$, and $\beta_{\rm ext}$.\footnote{For $\beta_{\rm ext} \gg 1$, the value of $r_{\rm ext,max}$ is of minor importance. Here we adopt $r_{\rm ext,max} = 30r_{\rm ext,min}$.}
Two of them can be robustly constrained from standard observational arguments --- $r_{\rm ext,min} = r_{\rm BLR(IR)} \propto L_{\rm d}^{1/2}$, and $\xi_{\rm ext} = \xi_{\rm BLR(IR)} \sim 0.1$ (specific values are provided for each case analyzed in Section \ref{sec_cases}).
Parameters $\alpha_{\rm ext,max}$ and $\beta_{\rm ext}$ determine the scale height and the radial stratification of the external radiation emitting region, respectively. These parameters are poorly understood, but the results are sensitive mainly to the former. In this work we assume that $\alpha_{\rm ext,max} = 45^\circ$ and $\beta_{\rm ext} = 4$.
In the case of planar geometry, with $\alpha_{\rm ext,max} \lesssim 10^\circ$ \citep{2012arXiv1209.2291T}, we would need to introduce an additional geometrical correction factor of order $\sim 0.1-0.2$ \citep{2013ApJ...779...68S}.

We now calculate the energy density of external radiation fields in the emitting region co-moving frame at the jet axis at distance $r$ from the SMBH (point A).
Consider an infinitesimal volume element ${\rm d}V = {\rm d}A\,{\rm d}r_{\rm ext}$ located within the adopted geometry at some $(r_{\rm ext},\alpha_{\rm ext})$ (point B).
The energy density of direct accretion disk radiation at point B is $u_{\rm d}(r_{\rm ext}) \simeq L_{\rm d}/(4\pi cr_{\rm ext}^2)$.
The luminosity of the radiation reprocessed by this volume element is ${\rm d}L_{\rm ext} = \xi(r_{\rm ext})u_{\rm d}(r_{\rm ext})c\,{\rm d}A$.
Its contribution to the co-moving energy density of external radiation at point A is ${\rm d}u_{\rm ext}' = \mathcal{D}_{\rm ext}^2\,{\rm d}L_{\rm ext}/(4\pi cd_{\rm ext}^2)$, where $\mathcal{D}_{\rm ext} = \Gamma(1+\beta\cos\theta_{\rm ext})$ is the Doppler factor of point B with respect to point A in the emitting region co-moving frame, $\tan\theta_{\rm ext} = r_{\rm ext}\cos\alpha_{\rm ext}/(r_{\rm ext}\sin\alpha_{\rm ext}-r)$ gives the zenithal angle of point B with respect to point A, and $d_{\rm ext}^2 = (r_{\rm ext}\cos\alpha_{\rm ext})^2 + (r-r_{\rm ext}\sin\alpha_{\rm ext})^2$ gives the distance between points A and B.
We also calculate the characteristic co-moving photon energy $E_{\rm ext}' = \mathcal{D}_{\rm ext}E_{\rm ext}$, where $E_{\rm ext}$ is independent of $(r_{\rm ext},\alpha_{\rm ext})$.
We integrate function $u_{\rm ext}'(E_{\rm ext}')$ over the entire volume of the adopted geometry, and we identify its peak value $u_{\rm ext,peak}'$ (in the $E u_E'$ sense), and the corresponding photon energy $E_{\rm ext,peak}'$.

Finally, we identify simple analytical forms that can reasonably well approximate the numerically calculated functions $u_{\rm ext}'(r)$ and $E_{\rm ext,peak}'(r)$. These forms are presented in Equations (\ref{eq_zeta}) and (\ref{eq_Eext}) in Section \ref{sec_constr_ssc}.

\section{Gamma-ray/mm connection}
\label{sec_disc_gamma_mm}

\begin{figure}[t]
\centering
\includegraphics[width=\columnwidth]{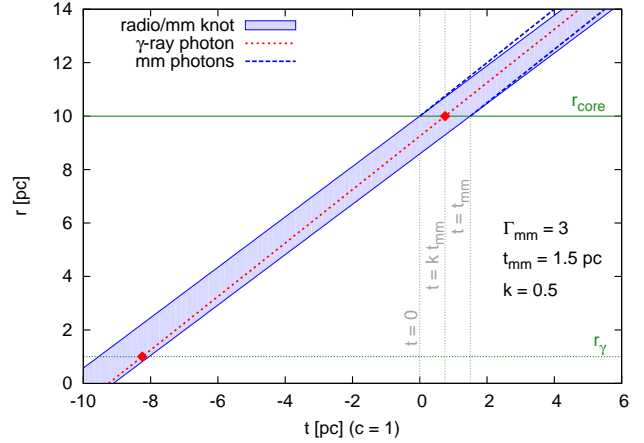}
\caption{Spacetime diagram illustrating the ambiguity of using the observational gamma-ray/mm connection to infer the location of the gamma-ray flares in blazars. In this example, we adopt exaggerated values of $\Gamma_{\rm mm}$ and $t_{\rm mm}$ to clearly distinguish the photons from the radio/mm knot. \emph{Red diamonds} indicate 2 events (out of many) consistent with the production of a gamma-ray flare within the radio/mm knot, but at widely different distances along the jet. See Appendix \ref{sec_disc_gamma_mm} for detailed description.}
\label{fig_gamma_mm}
\end{figure}

As we discussed in Section \ref{sec_ao0235}, the observational connection between many major gamma-ray flares and radio/mm activity of blazar jets has been used to argue that gamma-ray flares should be produced close to the location of radio/mm cores, at the distance scale $r_{\rm core} \simeq 10\;{\rm pc}$.
Here, we use a very simple light travel time argument to demonstrate that this inference is not valid.
The radio/mm activity typically consists of a $t_{\rm mm} \sim 100\;{\rm d}$ long radio/mm outburst and a superluminal radio/mm knot propagating downstream from the core, whose estimated moment of crossing the radio core coincides with the radio/mm outburst.
We approximate the superluminal knot by a shell of fixed thickness $l_{\rm mm}$ propagating with the Lorentz factor $\Gamma_{\rm mm} = (1-\beta_{\rm mm}^2)^{-1/2} \simeq 20$.
We relate the shell thickness to the radio/mm outburst duration by $l_{\rm mm} \simeq \beta_{\rm mm}ct_{\rm mm} \simeq 0.084\;{\rm pc}$.
We choose the time coordinate such that at $t = 0$ the front of the shell crosses the location of the radio/mm core, and thus the tail of the shell crosses the radio/mm core at $t = t_{\rm mm}$ (see Figure \ref{fig_gamma_mm}).
A gamma-ray flare is `observed' (gamma-ray photons cross the radio/mm core) at $t_{\rm\gamma,obs} = kt_{\rm mm}$, where $0 < k \lesssim 1$.
However, we assume that the gamma-ray flare was produced at $r_\gamma \simeq 1\;{\rm pc}$.
Thus, the gamma-ray photons were emitted at $t_{\rm\gamma,em} = t_{\rm\gamma,obs} - (r_{\rm core}-r_\gamma)/c$.
At that time, the front of the shell was located at $r_2 \simeq r_\gamma + (r_{\rm core}-r_\gamma)/(2\Gamma_{\rm mm}^2) + kl_{\rm mm}$, and its tail at $r_1 = r_2 - l_{\rm mm}$.
We can see that $r_2 > r_\gamma$, while the criterion for $r_1 < r_\gamma$, which means that the gamma-ray emission site was within the shell, is:
\be
k < 1 - \frac{r_{\rm core}-r_\gamma}{2\Gamma_{\rm mm}^2\beta_{\rm mm}ct_{\rm mm}}\,.
\ee
As long as $(r_{\rm core}-r_\gamma) \ll 2\Gamma_{\rm mm}^2\beta_{\rm mm}ct_{\rm mm}$, it is easy to have the gamma-ray flare produced within the shell.
For our fiducial parameters, this criterion is $k < 0.87$.
One can see that temporal coincidence, and even causality, between the gamma-ray flares and the radio/mm outburst does not imply that they are produced co-spatially.

\section{Far-dissipation solutions}
\label{sec_disc_far_diss}

We showed that two of our constraints, the $L_{\rm SSC}$ constraint and the $E_{\rm cool,obs}$ constraint, are likely violated at large distance scales. Here, we consider formal requirements to satisfy these constraints for an arbitrary $(r_0,\Gamma_0)$. From the $L_{\rm SSC}$ constraint (Equation \ref{eq_constr_LSSC}), we find the following condition:
\bea
u_{\rm ext}' &>& 0.09\;{\rm erg\,cm^{-3}}\times(1+z)^2
\left(\frac{\mathcal{D}}{\Gamma}\right)^{-8}
\left(\frac{\Gamma_0}{20}\right)^{-6}\times
\nonumber\\&&
\left(\frac{t_{\rm var,obs}}{1\;{\rm d}}\right)^{-2}
\left(\frac{L_{\gamma,48}L_{\rm syn,47}}{L_{\rm X,46}}\right)
\,.
\eea
And from the $E_{\rm cool,obs}$ constraint (Equation \ref{eq_constr_Ecool}), we find:
\bea
u_{\rm ext}' &>& 0.11\;{\rm erg\,cm^{-3}}\times(1+z)^{1/2}
\left(\frac{\mathcal{D}}{\Gamma}\right)^{-1/2}\times
\nonumber\\&&
\left(\frac{t_{\rm var,obs}}{1\;{\rm d}}\right)^{-1}
\left(\frac{E_{\rm ext}}{10\;{\rm eV}}\right)^{1/2}
\,.
\eea
Both the $L_{\rm SSC}$ and $E_{\rm cool,obs}$ constraints can be satisfied for a sufficiently high external radiation density.
For comparison, typical co-moving energy densities of BLR and IR components are of order $u_{\rm BLR}' \sim 15\;{\rm erg\,cm^{-3}}(\xi_{\rm BLR}/0.1)(\Gamma_0/20)^2$ for $r \lesssim 0.1\;{\rm pc}$ and $u_{\rm IR}' \sim 0.024\;{\rm erg\,cm^{-3}}(\xi_{\rm IR}/0.1)(\Gamma_0/20)^2$ for $r \lesssim 2.5\;{\rm pc}$, respectively.
The proposers of the far-dissipation scenarios have recognized the requirement for additional sources of external radiation.
In the following, we will evaluate two particular scenarios: a spine-sheath model \citep[\eg,][]{2010ApJ...710L.126M}, and an extended broad-line region \citep{2011A&A...532A.146L}.

\subsection{Spine-sheath models}
\label{sec_disc_spine_sheath}

In the spine-sheath model, the jet consists of a highly-relativistic spine surrounded by a mildly-relativistic sheath \citep{2005A&A...432..401G}.
Let us denote the spine co-moving frame with $\mathcal{O}'$, and the sheath co-moving frame with $\mathcal{O}''$.
Consider that the gamma-ray flares are produced in the spine by Comptonization of synchrotron radiation originating from the sheath, and that the synchrotron radiation from the spine region contributes significantly to the observed optical/IR emission.
The required energy density of the sheath radiation in $\mathcal{O}'$ is $u_{\rm sh}' \simeq 0.1\;{\rm erg\,cm^{-3}}$.
If the sheath propagates with Lorentz factor $\Gamma_{\rm sh}$ in the external frame, the radiation energy density in $\mathcal{O}''$ is $u_{\rm sh}'' \simeq u_{\rm sh}' / (4\Gamma_{\rm rel}^2/3)$, where $\Gamma_{\rm rel} = \Gamma_{\rm sh}\Gamma(1-\beta_{\rm sh}\beta) \simeq \Gamma/(2\Gamma_{\rm sh})$ is the relative Lorentz factor of $\mathcal{O}''$ in $\mathcal{O}'$ (the approximation is done in the limit where $1 \ll \Gamma_{\rm sh} \ll \Gamma$).
We can calculate the apparent luminosity of the sheath radiation for an external observer aligned with the jet spine as $L_{\rm sh,obs} \simeq 4\pi c\Gamma_{\rm sh}^4R_{\rm sh}^2u_{\rm sh}''$, where $R_{\rm sh} \simeq \theta_{\rm sh}r$ is the sheath radius parametrized by the sheath opening angle $\theta_{\rm sh}$.
Putting this all together, we find:
\bea
L_{\rm sh,obs}
&\simeq&
\frac{12\pi c\Gamma_{\rm sh}^4(\Gamma_{\rm sh}\theta_{\rm sh})^2r^2u_{\rm sh}'}{\Gamma^2} \simeq 2.7\times 10^{47}\;{\rm erg\,s^{-1}}\times
\nonumber\\&&
\Gamma_{\rm sh}^4(\Gamma_{\rm sh}\theta_{\rm sh})^2\left(\frac{r}{10\;{\rm pc}}\right)^2\left(\frac{\Gamma}{20}\right)^{-2}\,.
\eea
Assuming that $\Gamma_{\rm sh}\theta_{\rm sh} \sim 1$, even for very moderate values of $\Gamma_{\rm sh}$, we would have $L_{\rm sh,obs} > L_\gamma$, and even for higher values of $\Gamma$ it is very likely that $L_{\rm sh,obs} > L_{\rm syn}$.
The fact that $L_{\rm sh,obs}$ is a strongly increasing function of $\Gamma_{\rm sh}$ means that the spine-sheath model actually offers no advantage in providing soft photons for Comptonization to the observed gamma-ray emission over static sources of external radiation.

\subsection{Extended broad-line region}
\label{sec_disc_ext_blr}

Here we estimate a possible contribution to the external radiation energy density from a broad-line region extended along the jet to supra-pc distance scales \citep{2011A&A...532A.146L}.
For the purpose of first-order estimates, we will approximate the extended BLR as a sphere of radius $R_{\rm BLR*}$ centered on the jet at the distance scale $r_{\rm BLR*} \sim 10\,{\rm pc} \gg R_{\rm BLR*}$.
Let $L_{\rm BLR*}$ be the luminosity of emission lines produced in this region, not taking into account any lines produced elsewhere.
These emission lines are expected to be significantly narrower from conventional broad emission lines, and there is little observational evidence for their existence in the line profiles of radio-loud quasars.
Therefore, we will adopt an upper limit of $L_{\rm BLR*} \lesssim 10^{44}\;{\rm erg\,s^{-1}}$, so that it constitutes only a small fraction of the total luminosity of broad emission lines.
This luminosity will contribute the external radiation density $u'_{\rm BLR*} \simeq \Gamma^2L_{\rm BLR*}/(3\pi cR_{\rm BLR*}^2)$ at the center of the sphere in the co-moving frame of the gamma-ray emitting region.
We consider two sources of radiation illuminating the extended BLR --- 1) the direct accretion disk radiation of luminosity $L_{\rm d}$; and 2) the jet synchrotron radiation produced at an arbitrary distance scale $r_{\rm syn} < r_{\rm BLR*}$ and of apparent luminosity $L_{\rm syn}$.
We consider two types of covering factors --- the geometric factor $\xi_{\rm geom}$, and the intrinsic factor $\xi_{\rm int}$ --- such that $L_{\rm BLR*} = \xi_{\rm int}\xi_{\rm geom}L_{\rm d(syn)}$.

In case 1), assuming that the accretion disk radiation is roughly isotropic, the geometric factor is given by $\xi_{\rm geom} \simeq (R_{\rm BLR*}/2r_{\rm BLR*})^2$, hence:
\bea
u_{\rm BLR*}'
&\simeq&
\frac{\xi_{\rm int}\Gamma^2L_{\rm d}}{12\pi cr_{\rm BLR*}^2}
\simeq
3.7\times 10^{-4}\;{\rm erg\,cm^{-3}}\times
\nonumber\\&&
L_{\rm d,46}\left(\frac{\xi_{\rm int}}{0.1}\right)\left(\frac{\Gamma}{20}\right)^2\left(\frac{r_{\rm BLR*}}{10\;{\rm pc}}\right)^{-2}\,.
\eea
This value is more than 2 orders of magnitude too small to satisfy the $L_{\rm SSC}$ and $E_{\rm cool,obs}$ constraints for typical parameter values.

In case 2), the jet synchrotron radiation is strongly beamed, and effectively it can illuminate a region of radius $R_{\rm BLR*} \simeq (r_{\rm BLR*}-r_{\rm syn})/\Gamma$.
Assuming that all of the illuminated region is filled with the gas, we adopt $\xi_{\rm geom} \simeq 1$.\footnote{The synchrotron radiation beam should extend significantly beyond the jet boundaries (otherwise $\xi_{\rm geom} < 1$), which requires either that $\Gamma_{\rm j}\theta_{\rm j} \ll 1$, or that the jet accelerates significantly between $r_{\rm syn}$ and $r_{\rm BLR*}$. If the synchrotron radiation is produced in compact relativistic outflows called minijets, it can be pointed away from the jet cone with substantial probability \citep{2010MNRAS.402.1649G}.}
Since we normalize the synchrotron luminosity to $L_{\rm syn} \simeq 10^{47}\;{\rm erg\,s^{-1}}$, for consistency we adopt $\xi_{\rm int} \sim 10^{-3}$ in order to have $L_{\rm BLR*} \simeq \xi_{\rm int}L_{\rm syn} \sim 10^{44}\;{\rm erg\,s^{-1}}$:
\bea
u_{\rm BLR*}'
&\simeq&
\frac{\xi_{\rm int}\Gamma^4L_{\rm syn}}{3\pi c(r_{\rm BLR*}-r_{\rm syn})^2}
\simeq 0.24\;{\rm erg\,cm^{-3}}\times
\\&&
L_{\rm syn,47}\left(\frac{\xi_{\rm int}}{10^{-3}}\right)\left(\frac{\Gamma}{20}\right)^4\left(\frac{r_{\rm BLR*}-r_{\rm syn}}{5\;{\rm pc}}\right)^{-2}\,.
\nonumber
\eea
In principle, this mechanism can provide enough of external radiation density to satisfy the $L_{\rm SSC}$ and $E_{\rm cool,obs}$.
However, we suggest that more observational support for the existence of such emission lines is necessary.

\end{document}